\documentclass{jfm}
\usepackage{graphicx}
\usepackage{epstopdf, epsfig}

\usepackage[utf8]{inputenc}
\usepackage{amsmath}
\usepackage{amssymb}
\usepackage{mathrsfs}  
\usepackage{subcaption}
\usepackage[a4paper,margin = 3.1cm]{geometry}
\usepackage[UKenglish]{babel}
\usepackage{bm}
\usepackage{graphicx}
\usepackage{xcolor}
\usepackage{bigints}

\usepackage{sspcommands}
\usepackage{lineno}

\usepackage{graphicx}
\usepackage{hyperref}
\hypersetup{
    colorlinks=true,
    linkcolor=blue,
    citecolor =blue,
    filecolor=magenta,      
    urlcolor=cyan,
    pdftitle={Overleaf Example},
    pdfpagemode=FullScreen,
    }

\DeclareMathOperator{\Ra}{\text{Ra}}
\DeclareMathOperator{\Rs}{\text{Rs}}

\shorttitle{Salt fingering staircases and the three-component Phillips effect}
\shortauthor{P. Pru\v{z}ina, D. W. Hughes and S. S. Pegler}

\title{Salt fingering staircases and the three-component Phillips effect}

\author{Paul Pru\v{z}ina\aff{1}
  \corresp{\email{mmpep@leeds.ac.uk}},
  David W. Hughes\aff{1}
 \and Samuel S. Pegler\aff{1}}

\affiliation{\aff{1}School of Mathematics, University of Leeds, Leeds LS2 9JT, UK}

\begin{document}

\maketitle

\begin{abstract}
Understanding the dynamics of staircases in salt fingering convection presents a long-standing theoretical challenge to fluid dynamicists. Although there has been significant progress, particularly through numerical simulations, there are a number of conflicting theoretical explanations as to the driving mechanism underlying staircase formation. The Phillips effect proposes that layering in stirred stratified flow is due to an antidiffusive process, and it has been suggested that this mechanism may also be responsible for salt-fingering staircases. However, the details of this process, as well as mathematical models to predict the evolution and merger dynamics of staircases, have yet to be established. We generalise the theory of the Phillips effect to a  three-component system (e.g.\  temperature, salinity, energy) and demonstrate a regularised nonlinear  model  of layering based on mixing-length parameterisations. The model predicts both the inception of layering and its long-term  evolution through mergers, while generalising, and remaining consistent with, previous results for double-diffusive layering based on flux ratios. Our model of salt fingering 
is formulated using spatial averaging processes and closed by a mixing length parameterised in terms of the kinetic energy and the ratio of the temperature and salt gradients. The model predicts a layering instability for a bounded range of parameter values in the salt fingering regime. Nonlinear solutions show that an initially unstable linear buoyancy gradient develops into layers, which proceed to merge through a process of stronger interfaces growing at the expense of weaker ones. 
Our results indicate that these mergers are responsible for the characteristic increase of buoyancy flux through thermohaline  staircases.
\end{abstract}


\section{Introduction}
The dynamics of thermohaline staircases in double-diffusive systems has interested oceanographers since they were first reported in the 1960s. Oceanographic observations, laboratory experiments and direct numerical simulations have all shown the existence of wide, evenly mixed layers separated by sharp interfaces, lasting for long times---on the order of months---with little change.  
Thermohaline convection is a specific example of double-diffusive convection (DDC), in which motions are driven in a stably stratified fluid as a result of the buoyancy depending on two independent scalar quantities that diffuse at different rates. In accordance with the oceanic (thermohaline) context, we refer to the faster diffusing component as `temperature' and the slower as `salinity'. DDC occurs in two different regimes, depending on which component of buoyancy provides the destabilising gradient. \textit{Salt fingering} refers to a configuration in which the temperature gradient is stabilising and the salinity gradient destabilising. \textit{Diffusive convection}, on the other hand, occurs when the temperature gradient is destabilising and the salinity gradient stabilising. In the oceans, staircases have been found both in regions susceptible to salt fingering \citep{TaitHowe, Schmitt_et_al_2005} and diffusive convection \citep{Nealetal, Timmermans_et_al_2008}; their existence is also widely documented, in both regimes, in numerical studies \citep[e.g.][]{Radko_2003, Stellmach_et_al_2011, Rosenblum_et_al_2011, HB_2021}. Despite their prevalence, the conditions for the development of staircases, as well as the dynamics of their long-term evolution and merger, are not well understood. 

Several theories have been proposed for the driving mechanism behind layering, which are well documented in reviews by \citet{Merryfield} and \citet{radko_2013}. An early hypothesis was that of collective instability \citep{stern_1969}, in which growing salt fingers excite large-scale internal waves that overturn and generate a stepped structure.  It has also been suggested that staircases are the long-time state of thermohaline intrusions \citep{ZhurbasOzmidov,Merryfield}, or that they are metastable equilibria of the system, requiring a finite amplitude perturbation from an initial linearly stable state \citep{Veronis1965, Stern_Turner_1969}. Other models rely on heating a stably stratified fluid from below, with convective layers forming sequentially from the bottom upwards \citep{turner_stommel_1964, huppert_linden_1979}.

 A further idea is that of \citet{Radko_2003}, who proposed that the driving factor behind staircases is the result of an instability arising from 
 variation of the ratio of the thermal to solutal fluxes.  With $T(z,t)$ and $S(z,t)$ representing the horizontally-averaged temperature and salinity fields, and subscripts of $t$ and $z$ denoting partial derivatives with respect to time and height, \citet{Radko_2003} models the two components contributing to the density by
\begin{equation}
    T_t = f_z,\hspace{5mm} S_t = c_z,
    \label{eqn:Radkomodel}
\end{equation}
where $f(R)$ is the temperature flux and $c(R)$ the salinity flux, dependent on the density ratio $R = T_z/S_z$; the flux ratio $\gamma(R)$ is defined by  $\gamma = f/c$. The growth rate of perturbations is found to be positive if and only if $\ttd{\gamma}{R}$ is negative. On the basis of  numerical simulations, \cite{Radko_2003} argues that $\gamma(R)$ should be non-monotonic with a single minimum, so that there is an unstable range of $R$, but the instability is arrested in regions where $\ttd{\gamma}{R} > 0$. 
Radko's model provides a helpful conceptual framework for relating  the condition for instability in terms of properties of the buoyancy fluxes. However, it describes only the conditions for an initial linear instability, with a growth rate that diverges at infinite wavenumber. As identified by \citet{radko2019thermohaline}, such an ultraviolet catastrophe precludes the identification of a preferred wavelength or maximal growth rate, and  prevents  its use to study the dynamics of larger scale layers.
A more complex regularised physical theory is required to model the full evolution from initial perturbation to staircase.  In addition, the flow velocity is absent from the model, with the fluxes depending only on the density ratio $R$. 
To regularise the high wavenumber instability, \citet{radko2019thermohaline} proposed a model based on an asymptotic multiscale analysis, which leads to hyperdiffusion terms in the temperature and salinity equations, giving negative growth rates at high wavenumbers. Thus the flux-gradient model can be adapted to study the evolution beyond an initial instability to a large scale staircase structure.

The formation of layers is antidiffusive, with up-gradient transport, representing behaviour contrary to the  homogenisation one might expect. Early work on layering by \citet{PHILLIPS197279} and \citet{Posmentier} proposed a mechanism for the development of staircases in a stirred stratified fluid with a single component of buoyancy. They appealed directly to this antidiffusive property, based on the turbulent diffusion of buoyancy with a diffusion coefficient that could be negative. The buoyancy field is modelled by a one-dimensional turbulent diffusion equation
\begin{equation}
b_t = \pd{}{ z}f(b_z),\label{eqn:Phillipseqn}
\end{equation}
where $f(b_z)$ is a flux function dependent on the buoyancy gradient $b_z$. A linear stability analysis shows that a uniform buoyancy gradient $b_z = g_0$ is unstable to perturbations if
\begin{equation}
\td{f}{b_z}(g_0) < 0.\label{eqn:Phillipscond}
\end{equation}
If condition \eqref{eqn:Phillipscond} holds for a finite range of $g_0$, then, in this unstable range, perturbations will grow. However, this is a prediction only from linear theory and, furthermore, leads to an ill-posed high-wavenumber instability with a growth rate of  $s\sim m^2$ as wavenumber $m\to\infty$. Nonetheless, \citet{Posmentier} proposed that perturbations do not remain unstable indefinitely, but stabilise once the gradient reaches a value outside the unstable region, evolving into a stepped structure.

To regularise the dynamics, \citet{BLY98} (hereinafter BLY) coupled the buoyancy equation \eqref{eqn:Phillipseqn} to an energy equation, considering the system
\begin{eqnarray}
g_t &=& f_{zz},\label{eqn:BLYg}\\
e_t &=& \left(\kappa e_z\right)_z + p,\label{eqn:BLYe}
\end{eqnarray}
where $g(z,t)$ is the buoyancy gradient, $f(z,t)$ is some flux function, $e(z,t)$ is the turbulent kinetic energy, $\kappa$ is a turbulent diffusion coefficient and $p$ is a general source (production) of energy. In the absence of double-diffusive effects, a parameterisation for $p$ must include an energy source to drive the layering process. For linear instability in the system~\eqref{eqn:BLYg}--\eqref{eqn:BLYe}, the equivalent of the Phillips condition~\eqref{eqn:Phillipscond} is
\begin{equation}
\td{f}{g}\equiv\frac{f_gp_e-f_ep_g}{p_e} < 0.\label{eqn:BLYcond}
\end{equation}
The high-wavenumber instability inherent to the Phillips model~\eqref{eqn:Phillipseqn} is avoided by parameterisations such that $\ttd{f}{g} < 0$ but $\partial f/\partial g > 0$. With this regularisation, \eqref{eqn:BLYg}--\eqref{eqn:BLYe} provide a complete model that can be used to analyse layer formation, evolution and merger in a stirred stratified fluid.
A crucial aspect to the model is the dependence of the turbulent fluxes on a turbulent mixing length $l(g,e)$, used to close the system.

Models of the general form~\eqref{eqn:BLYg}--\eqref{eqn:BLYe} have been used to study layering in several contexts. \citet{MD_2019} produced a similar style of model to describe the formation of potential vorticity staircases. \citet{Pruzina2022stirred} extended the BLY model in its original context of stirred stratified layering to examine the influence of the boundary conditions on the layering process and long-term merger trends. We will refer to these as \textit{two-component models}. However, to study double-diffusive convection, the buoyancy field must be split into two independent components (e.g. temperature and salt). Hence, to produce a BLY-style model for double-diffusion, a third equation must be added to account for the second component of buoyancy. 
\citet{PvH} adapted the BLY model to a double-diffusive context by arguing, on the basis of their numerical simulations~\citep{PvH_2012}, that the flux ratio $\gamma$ remained constant, and hence that the evolution of the temperature and salinity fields could be investigated with a single equation for the total buoyancy. This assumption reduces their model to a two-component model similar to that of BLY. Motivated by the results of numerical simulations \citep{PvH_2012}, the model of \citet{PvH} is forced by a constant up-gradient salt finger flux, with an eddy diffusivity term representing stirring due to `clusters' of salt fingers. The constant salt finger flux has no effect on the buoyancy equation, but contributes a positive source in the energy equation. As such, while the underlying physics is different, the model of \citet{PvH} takes a very similar form to that of BLY, modelling salt fingering staircases with a forced stratified system with a single component of buoyancy. The model of \citet{PvH} was studied theoretically by \citet{coclite}, who proved the existence of solutions, and discussed some of their properties.

However, it is useful to investigate whether layering is also possible without such forcing, instead including double-diffusion directly. In this case, a third equation is necessary that allows temperature and salinity to be described individually. There has been some limited use of three-component models to study ${\bm E} \times {\bm B}$ staircases in plasma drift-wave turbulence  \citep{AD_2016, AD_2017, Guo_et_al_2019}. In these 
systems, the instability takes place in only two of the equations, so the modelling of instability again reduces to a form similar to the  two-component BLY framework.

For a uniform buoyancy gradient to develop into a more complex layered structure, some energy input is necessary. BLY included an explicit source term to represent stirring, while \cite{PvH} included a constant background salt-finger flux. However, several computational studies of DDC have shown that no such external energy input is necessary for staircases to form. Instead, the double-diffusive instability provides a mechanism for the transfer of potential energy into kinetic energy. This is true in both the salt fingering regime \citep[e.g.][]{Stellmach_et_al_2011} and the diffusive convection regime \citep[e.g.][]{Rosenblum_et_al_2011,HB_2021}. As such, we seek to formulate a model with no prescribed external forcing.

\citet{ma2022thermohaline} note that the values of the density ratio $R$ found in observed oceanic staircases in the diffusive convection regime ($2<1/R<7$) differ significantly from the values predicted by classical linear stability theory ($1<1/R<1.14$) \citep[e.g.][]{turner_1973}. By contrast, in the salt fingering regime, the values of $R$ in observed staircases match well with linear theory. On this basis, \citet{ma2022thermohaline} suggest that the instability in diffusive convection is not actually the driver behind diffusive staircases, instead proposing that the layering instability relies on external forcing from a background flow, with double-diffusive effects being important only in the regularisation and stabilisation of layers. 

In the current work, we develop the first three-component turbulent mixing-length model of double-diffusive layering in terms of temperature, salinity and energy. We demonstrate its application to the analysis of layer evolution and merger dynamics. 

Our starting point is the stability analysis for a general three-component system in terms of two independent components of density and the turbulent kinetic energy. We categorise the  different modes of instability that can occur, and make comparisons with the previous theories of Phillips and BLY. The $\gamma$-instability of \citet{Radko_2003} is shown to be mathematically equivalent to the Phillips instability, in the case where the flux functions are parameterised in terms of the background density ratio $R = T_z/S_z$.

We present a new model for thermohaline staircases in turbulent flow, which is  derived from the Boussinesq equations using a horizontal averaging process. We apply this model  to the salt fingering regime of double-diffusive convection. We analyse the linear stability of steady states, and demonstrate that the system is susceptible to the Phillips instability for a range of parameter values within the salt fingering regime. Numerical solutions of the model to long times indicate that staircases evolve via the `B-merger' pattern described by \citet{radko2007mechanics}, whereby strong interfaces grow at the expense of weaker ones.  Each layer merger causes the buoyancy gradient in surviving interfaces to increase, and the buoyancy flux through the layers to increase, so the staircase has significantly higher buoyancy flux than the initial unlayered state.

The paper is arranged as follows. In \S\,\ref{sec:3Phillips}, we present a linear stability analysis of a general three-component system, thereby giving a general characteristic equation for growth rates in terms of vertical wavenumber. We analyse the solutions to the characteristic equation in the limit of small wavenumber in \S\,\ref{sec:smallwavenumber}, the limit of large wavenumber in \S\,\ref{sec:largewavenumber}, and the conditions for marginal stability in \S\,\ref{sec:marginalstability}. We discuss the conditions for layering in \S\,\ref{sec:layeringconditions}. In \S\,\ref{sec:Radko} we demonstrate the relationship between Radko's $\gamma$-instability and the Phillips effect.
Section~\ref{sec:model} contains a description of our three-component double-diffusive model and a discussion of the mixing length on which it depends. In \S\,\ref{sec:stability} we apply the results of \S\,\ref{sec:3Phillips} to our model, demonstrating the expected regions of instability, and discussing the effect of changing the parameters of the model. In \S\,\ref{sec:numerics}, we present long-term numerical solutions and discuss the behaviour of the buoyancy flux through layer mergers. We end in \S\,\ref{sec:conc} by summarising our key conclusions.

\section{The three-component Phillips effect}\label{sec:3Phillips}
To develop a theory for three-component models, we first investigate the linear stability properties of a general three-component system of the form
\begin{eqnarray}
g_t &=& f_{zz},\label{eqn:generalg}\\
d_t &=& c_{zz},\label{eqn:generald}\\
e_t &=& \left(\kappa e_z\right)_z + p.\label{eqn:generale}
\end{eqnarray}
Here, $g(z,t)$ and $d(z,t)$ are the independent components of the buoyancy gradient, with $f(g,d,e)$ and $c(g,d,e)$ their corresponding turbulent fluxes. To analyse the conditions for instability in this general system, we perform a linear stability analysis. We assume that $(g_0,d_0,e_0)$ is a uniform steady state, such that $p(g_0,d_0,e_0) = 0$ and that $(g',d',e')$ is a small perturbation. From the chain rule, we can write the $z$-derivative as
\begin{equation}
\frac{\partial}{\partial z} =\frac{\partial g}{\partial z}\frac{\partial}{\partial g} + \frac{\partial d}{\partial z}\frac{\partial}{\partial d}  + \frac{\partial e}{\partial z}\frac{\partial}{\partial e}.
\label{eqn:zderivative}
\end{equation}
 On applying~\eqref{eqn:zderivative}, expanding $p(g,d,e)$ as a Taylor series, and  neglecting  terms quadratic in the perturbation quantities, we obtain the linear form of the general model:
\begin{align}
g'_t &\approx g'_{zz}f_g + d'_{zz}f_d + e'_{zz}f_e,\\
d'_t &\approx g'_{zz}c_g + d'_{zz}c_d + e'_{zz}c_e,\\
e'_t &\approx e'_{zz}\kappa  + g'p_g + d'p_d + e'p_e,
\end{align}
where the partial derivatives $f_g$ etc. are evaluated in the uniform steady state. On seeking solutions of the form $(g',d',e')\propto\exp(st+imz)$, where $s$ is the growth rate and $m$ the vertical wavenumber, the linearised forms of  \eqref{eqn:generalg}--\eqref{eqn:generale} may be expressed in matrix form as
\begin{equation}
\begin{pmatrix} s+m^2 f_g & m^2 f_d & m^2 f_e \\ m^2 c_g & s+m^2 c_d & m^2 c_e \\ -p_g & -p_d & s+m^2\kappa - p_e \end{pmatrix}\begin{pmatrix}g_1\\d_1\\e_1\end{pmatrix} = 0. \label{eqn:matsyst}
\end{equation}
Equation~\eqref{eqn:matsyst} has a non-trivial solution only if the  determinant of the matrix is zero, leading to the characteristic equation
\begin{align}
  s^3 &+ s^2\big[ m^2(f_g + c_d + \kappa) - p_e \big]\nonumber\\ &+ s\big[ m^4(f_gc_d - f_dc_g + \kappa f_g + \kappa c_d) + m^2(f_ep_g - f_gp_e + c_ep_d - c_dp_e)\big]\nonumber\\ &+ m^6\kappa(f_gc_d - f_dc_g) + m^4(f_gc_ep_d - f_gc_dp_e + f_ec_dp_g - f_ec_gp_d + f_dc_gp_e - f_dc_ep_g)=0,
  \label{eqn:evaleqn}
\end{align}
forming a cubic equation relating the growth rate to the vertical wavenumber.

 We now explore the conditions for the existence of unstable wavenumbers ($\Re(s) > 0$). To gain an analytic foothold, we consider the asymptotic limits of small and large wavenumber $m$, noting that a positive growth rate $\Re(s) > 0$ in either limit is a sufficient condition for instability.

\subsection{Instability at small wavenumbers}\label{sec:smallwavenumber}
When $m = 0$, corresponding to infinitely long spatial scales, the characteristic equation~\eqref{eqn:evaleqn} reduces to
\begin{equation}
s^3 - p_es^2 = 0,\label{eqn:m=0}
\end{equation}
giving one root $s=p_e$ and two zero roots. If
\begin{equation}
-p_e<0,
\label{eqn:generalcond1}
\end{equation} 
there is growth in the energy equation alone, without requiring interaction from the temperature equations. This \textit{energy mode} instability was also theoretically possible in the two-component BLY formulation, but the parameterisation adopted by BLY produces $-p_e>0$ everywhere.

To determine the stability of the two zero roots of \eqref{eqn:m=0}, it is necessary to include higher order terms. On taking the limit $m \to 0$, the dominant balance in \eqref{eqn:evaleqn} results from $s=O(m^2)$, giving
\begin{equation}
s^2 + \left(F_g+C_d\right)m^2 s + \left(F_gC_d-F_dC_g\right)m^4 = 0,\label{eqn:smallm}
\end{equation}
where we have adopted the following notations for simplicity:
\begin{align}
F_g\equiv \td{f}{g}&=\frac{f_gp_e - f_ep_g}{p_e}, \label{eqn:Fgdef}\\
C_d\equiv\td{c}{d}&=\frac{c_dp_e - c_ep_d}{p_e}, \label{eqn:Cddef}\\
F_d\equiv \td{f}{d}&=\frac{f_dp_e - f_ep_d}{p_e}, \label{eqn:Fddef}\\
C_g \equiv \td{c}{g}&=\frac{c_gp_e - c_ep_g}{p_e}. \label{eqn:Cgdef}
\end{align}
Equation~\eqref{eqn:smallm} has at least one root with positive real part if either
\begin{subeqnarray}
\gdef\thesubequation{\theequation \textit{a,b}}
F_gC_d-F_dC_g<0\label{eqn:generalcond2}\qquad \text{or} \qquad F_g+C_d<0.
\end{subeqnarray}
If \eqrefl{eqn:generalcond2}{a} is satisfied, then there is exactly one positive root, implying that the state is unstable. If \eqrefl{eqn:generalcond2}{b} is satisfied, but not \eqrefl{eqn:generalcond2}{a}, then there are two positive roots. Condition~\eqrefl{eqn:generalcond2}{a} represents the direct equivalent of the Phillips effect in a three-component system; \eqrefl{eqn:generalcond2}{b} extends this to allow for an oscillatory instability.

These conditions can also be interpreted in a vector framework, which is helpful for generalising to an $N$-component system. Let $\bm{F}$ be the vector function
\begin{equation}
\bm{F}(\bm{G}) = \begin{pmatrix}f\left(g,d,e(g,d)\right)\\c\left(g,d,e(g,d)\right)\end{pmatrix},
\end{equation}
where $e(g,d)$ is defined implicitly via $p=0$. The Jacobian of $\bm{F}$ with respect to $\bm{G} = (g,d)$ is then
\begin{equation}
\bm{J} = \begin{pmatrix}F_g&F_d\\ C_g& C_d\end{pmatrix}.\label{eqn:jacobian}
\end{equation}
Hence, conditions~\eqrefl{eqn:generalcond2}{a,b} can be rewritten respectively as
\begin{equation}
\det(\bm{J}) < 0, \qquad \text{tr}(\bm{J}) < 0. \label{eqn:determinantcond}
\end{equation}
Together, these conditions are equivalent to the single condition that there is instability if $\bm{J}$ has at least one negative eigenvalue. The same condition can be obtained by considering the system in the general form
\begin{equation}
\pd{\mathbf{G}}{t} = \pdn{}{z}{2}  \bm{F}\left(\bm{G},e(\bm{G})\right),  \qquad p\left(\bm{G},e(\bm{G})\right) = 0.\label{eqn:Ndimgeneral}
\end{equation}
Equation~\eqref{eqn:Ndimgeneral} could be readily extended into a general $N$-dimensional system, producing instability if the Jacobian of $\bm{F}$ with respect to $\bm{G}$ has at least one negative eigenvalue.

\subsection{Instability at high wavenumbers}\label{sec:largewavenumber}
For $m \to \infty$, the characteristic equation~\eqref{eqn:evaleqn} simplifies at leading order to
\begin{equation}
  s^3 + s^2 m^2(f_g + c_d + \kappa)  + s m^4(f_gc_d - f_dc_g + \kappa f_g + \kappa c_d) + m^6\kappa(f_gc_d - f_dc_g)=0.
  \label{eqn:evaleqn2}
\end{equation}
In this limit, all three solutions obey $s = O(m^2)$.
There is at least one root $s$ with positive real part if either the $s$-independent term $m^6\kappa(f_gc_d-f_dc_g)$ is negative, or the characteristic equation has a stationary point with $s>0$. Assuming that $f_g$ $c_d$ and $\kappa$ are all positive, in order to avoid the high-wavenumber instability of \citet{PHILLIPS197279}, both of these conditions reduce to
\begin{equation}
f_g c_d - f_d c_g < 0,
\label{eqn:generalcond4}
\end{equation}
thereby providing us with a criterion for the existence of an unstable large wavenumber. Note that if there is no energy equation, then $p\equiv 0$ and $F_g = f_g$, etc. In this special case, conditions~\eqrefl{eqn:generalcond2}{a} and~\eqref{eqn:generalcond4} are identical, and the growth rate of the Phillips instability $s\to \infty$ as $m\to\infty$. This demonstrates how the inclusion of the energy equation~\eqref{eqn:generale} regularises the instability at high wavenumbers.

We note that if \eqref{eqn:generalcond4} is satisfied and there is a high wavenumber instability, then one option to regularise it is by the addition of hyperdiffusion terms. It is straightforward to show that adding $-Ag_{zzzz}$ and $-B d_{zzzz}$ to \eqref{eqn:generalg} and \eqref{eqn:generald} gives growth rates $s\sim-m^4$ as $m\to\infty$. For brevity, we omit the details here, as the parameterisations we will use do not lead to a high wavenumber instability.

\subsection{Condition for marginal stability}\label{sec:marginalstability}
To check for instability at intermediate wavenumbers, we consider the point of marginal stability. Setting $s=0$ in~\eqref{eqn:evaleqn}, we find the wavenumbers for marginal stability to be $m=0$ and 
\begin{equation}
m = m_* = \sqrt{\frac{p_e\left(F_gC_d-F_dC_g\right)}{\kappa\left(f_gc_d-f_dc_g\right)}}.
\end{equation}
If $m_*$ is real, then $s$ is of one sign for  $0<m<m_*$, and the opposite sign for $m>m_*$. There is only one positive value for $m_*$, so as $m$ is varied, $s$ can change sign no more than once. Likewise, the critical wavenumber for oscillatory instability can be found by setting $\Re(s) = 0$ in~\eqref{eqn:evaleqn}. In this case, $m_*$ is given by solutions to
\begin{align}
    m^4&\left(f_g+c_d\right)\left(f_gc_d-f_dc_g + \kappa\left(f_g+c_d+\kappa\right)\right)\nonumber\\ &+ m^2\left(\frac{1}{-p_e}\kappa\left(f_g+c_d + F_g+C_d\right) + \left(f_g+c_d\right)^2 + \frac{1}{-p_e}\left(f_gf_ep_g+c_dc_ep_d+f_dc_ep_g+f_ec_gp_d\right)\right)\nonumber \\ &+ (-p_e)\left(F_gC_d-F_dC_g\right) = 0.
\end{align}
Assuming that $f_g>0$ and $c_d>0$, in order to avoid the high-wavenumber instability of \citet{PHILLIPS197279}, the only possibility for a positive real solution $m_*$ to exist is if one of the previous instability conditions~\eqref{eqn:generalcond1}, \eqref{eqn:generalcond2}, \eqref{eqn:generalcond4} is satisfied.

Hence, the only possible instabilities are via condition~\eqref{eqn:generalcond1} (energy mode, $m\to0$, $m_*$ real), conditions~\eqrefl{eqn:generalcond2}{a,b} (small wavenumber Phillips instability, $m_*$ real), or condition~\eqref{eqn:generalcond4} (high wavenumber, $m_*$ real). A combination of the conditions is possible, such that $m_*$ is imaginary and there is a positive growth rate $\Re(s)>0$ for all $m$.

\subsection{Conditions for layering}\label{sec:layeringconditions}
\begin{figure}
\centering
\includegraphics[width = 0.84\textwidth]{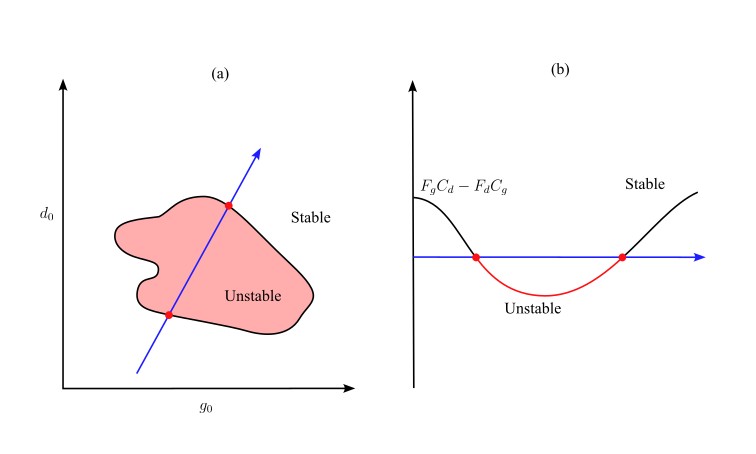}
\caption{($a$) Sketch of a region of instability in $g_0$-$d_0$ space, shaded pink. The locus of marginal stability is shown in black.  The blue line shows an arbitrary cross-section through the unstable region. Panel ($b$) shows the value of $F_gC_d-F_dC_g$ along the blue path --- it is negative only in the finite region between the two points shown in red, giving only a finite region where condition~\eqrefl{eqn:generalcond2}{a} is satisfied}
\label{fig:3eqnstab}
\end{figure}

We have deduced the conditions for linear instability, but have not yet demonstrated how these conditions can lead to layering. BLY proposed that, in a two-component model, layering requires an $N$-shaped relation between the buoyancy flux $f$ and the buoyancy gradient $g$, so that $F_g<0$  for only a finite range of gradients. The equivalent condition for a three-component model with independent contributions to the buoyancy is that there exits a bounded region of $g$-$d$ space in which any one of the conditions for instability \eqref{eqn:generalcond1}, \eqref{eqn:generalcond2} or \eqref{eqn:generalcond4} are met, as illustrated  schematically in figure~\ref{fig:3eqnstab}. On any path through the unstable region, only a finite range of points are unstable, with stable regions either side arresting the instability.

\subsection{Comparison with Radko's \texorpdfstring{$\gamma$}{gamma}-instability}\label{sec:Radko}
\citet{Radko_2003} put forward the idea that the driving factor behind layering is an instability arising from the parametric variation of the flux ratio $\gamma$ as a function of the density ratio $R$. He modelled the two components of the density as
\begin{eqnarray}
T_t &=& \frac{\partial}{\partial z}f(\gamma,Nu),\label{eqn:RadkoT}\\
S_t &=& \frac{\partial}{\partial z}c(\gamma,Nu),\label{eqn:RadkoS}
\end{eqnarray}
where the flux ratio $\gamma = f/c$ and the Nusselt number $N\!u$ is the ratio of convective to conductive heat transfer. The functions $\gamma(R)$ and $N\!u(R)$ depend only on the density ratio $R = \alpha T_z/\beta S_z$, i.e.\ the ratio of the contributions to the density from temperature and salt. Steady states of the flux-gradient relations~\eqref{eqn:RadkoT}--\eqref{eqn:RadkoS} are found to be linearly unstable to perturbations when 
\begin{equation}
\frac{\mathrm{d}\gamma}{\mathrm{d}R}<0, \label{eqn:gamma}
\end{equation}
representing the $\gamma$-instability.

To compare the criterion~\eqref{eqn:gamma} with the conditions for instability that we have deduced in the context of our general three-component model, we formulate our system in terms of Radko's parameters. 
Making Radko's assumption that the equations can be parameterised in terms only of the ratio of gradients, $R$, we express the fluxes in terms of turbulent diffusivities $K_T(R)$ and $K_S(R)$ as
\begin{eqnarray}
f &=& K_T(R)g,\label{eqn:Radkotflux}\\
c &=& K_S(R)d,\label{eqn:Radkosflux}
\end{eqnarray}
with $R = g/d$. Returning to the Jacobian form of the problem~\eqref{eqn:jacobian} with this new notation, we obtain
\begin{equation}
\bm{J} = \begin{pmatrix}K_T + K_T'R&-K_T'R^2\\K_S'&K_S - K_S'R\end{pmatrix},
\end{equation}
where primes denote differentiation with respect to $R$. Note that $\bm{J}$ does not depend on $g$ or $d$ independently and the only parameter of importance is the density ratio $R$. The trace and determinant of $\bm{J}$ are given by
\begin{eqnarray}
\text{tr}(\bm{J}) &=& K_T + K_S + R(K_T'-K_S'),\\
\det(\bm{J}) &=& K_TK_S + R(K_T'K_S - K_TK_S').
\end{eqnarray}
With fluxes of the form~\eqref{eqn:Radkotflux}--\eqref{eqn:Radkosflux}, the flux ratio $\gamma = RK_T/K_S$, so
\begin{equation}
\frac{\mathrm{d}\gamma}{\mathrm{d}R} = \frac{\mathrm{d}}{\mathrm{d}R}\left(R\frac{K_T}{K_S}\right) =  \frac{K_T}{K_S}   + \frac{R}{K_S^2}\left(K_T'K_S-K_TK_S'\right) = K_S^2\mathrm{det}(\bm{J}).
\end{equation}
The negative determinant condition~\eqref{eqn:determinantcond} exactly recovers Radko's $\gamma$-condition. The trace condition~\eqrefl{eqn:determinantcond}{b} is new, and allows for two unstable modes.

To summarise, Radko's $\gamma$-condition \eqref{eqn:gamma} is mathematically equivalent to the Phillips instability, in the specific context of double-diffusive flux-gradient relations dependent on $R$. The three-component model~\eqref{eqn:generalg}--\eqref{eqn:generale}, with instability conditions~\eqref{eqn:generalcond1}, \eqref{eqn:generalcond2} and \eqref{eqn:generalcond4}, describes a generalisation of both the Phillips and $\gamma$-instabilities for a three-component system with explicit dependence on the kinetic energy $e$. The inclusion of $e$ avoids the ultraviolet catastrophe inherent to the $\gamma$-instability and the single-component Phillips instability, allowing the model to capture not only the initial growth of perturbations but also the possible development of layers and their long-term evolution. Note that for a system in the general form~\eqref{eqn:generalg}--\eqref{eqn:generale},  condition~\eqrefl{eqn:generalcond2}{a} can lead to instability by two different physical mechanisms. If the function $p$ is parameterised to include an energy source term, then the system describes the forced mechanism of BLY and \citet{PvH}. By contrast, with no source term in $p$, but appropriate parameterisations for $f$ and $c$ such that \eqrefl{eqn:generalcond2}{a} is satisfied, the instability comes from the $\gamma$-style instability analagous to that of \citet{Radko_2003}.

\section{A three component model for thermohaline staircases}\label{sec:model}

To develop our model for the formation and evolution of thermohaline staircases, we consider a domain of height $h$, with a background dimensional temperature gradient $\Theta_z$,  salinity gradient $\Sigma_z$ and reference density $\rho_0$. The evolution of the velocity $\bm{u}(\xb,t)$, temperature $T(\xb,t)$ and salinity $S(\xb,t)$ are governed by the Boussinesq equations
\begin{eqnarray}
\bm{u}_{t} + \bm{u}\cdotb\nablab\bm{u} &=& -\frac{1}{\rho_0}\nablab p +  \frac{g\left(\rho-\rho_0\right)}{\rho_0} \eb_z + \nu\nabla^2\bm{u},\label{eqn:boussu}\\
T_{t} + \bm{u}\cdotb\nablab T &=& \kappa_T \nabla^2 T,\\
S_{t} + \bm{u}\cdotb\nablab S &=& \kappa_S \nabla^2 S,\\
\nablab\cdotb\bm{u}&=&0,\\
\frac{\rho-\rho_0}{\rho_0} &=& \beta S - \alpha T,
\label{eqn:boussstate}
\end{eqnarray}
where ${\rho}(\xb,t)$ is the density, and $p(\xb, t)$ the pressure. The equations depend on the kinematic viscosity $\nu$, the thermal and solutal diffusivities $\kappa_T$ and $\kappa_S$, gravitational acceleration $g$, and thermal and solutal expansion coefficients $\alpha$ and $\beta$. 

We nondimensionalise the system~\eqref{eqn:boussu}--\eqref{eqn:boussstate} via
\begin{equation}
\hat{t} = \frac{\kappa_T}{L^2}t,\quad \hat{z} = \frac{1}{L},\quad \hat{\bm{u}} = \frac{L}{\kappa_T}\bm{u},\quad \hat{T} = \frac{\alpha gL^3}{\kappa_T\nu}T,\quad \hat{S} = \frac{\beta gL^3}{\kappa_T\nu}S,\quad \hat{p} = \frac{L^2}{\rho_0 \nu \kappa_T }p,\label{eqn:ddcnondim}
\end{equation}
with hats denoting dimensionless quantities. The characteristic length $L$ is taken to be the salt finger length, chosen such that the magnitude of the local Rayleigh number is equal to unity \citep{Stern1960}:
\begin{equation}
    |\Ra| = \frac{\alpha g|\Theta_z|L^4}{\kappa_T\nu} = 1.
    \end{equation}
With the choice of nondimensionalisation~\eqref{eqn:ddcnondim}, the magnitudes of the dimensionless background temperature and salinity gradients are equal to the thermal and solutal Rayleigh numbers:
\begin{align}
|\hat{\Theta}_z| &=  \frac{\alpha g |\Theta_z|L^4}{\kappa_T\nu} = |\Ra| = 1,\\
|\hat{\Sigma}_z| &= \frac{\beta g |\Sigma_z|L^4}{\kappa_T\nu} = |\Rs| = \frac{1}{R_0},
\end{align}
where $R_0$ is the density ratio. Dropping hats, we obtain the dimensionless form of the governing equations~\eqref{eqn:boussu}--\eqref{eqn:boussstate} as
\begin{eqnarray}
\bm{u}_t + \bm{u}\cdotb\nablab\bm{u} &=& - \sigma\nablab p + \sigma b \eb_z + \sigma \nabla^2\bm{u}, \label{eqn:boussunondim}\\
T_t + \bm{u}\cdotb\nablab T &=& \nabla^2T,\label{eqn:boussTnondim}\\
S_t + \bm{u}\cdotb\nablab S &=& \tau \nabla^2 S,\label{eqn:boussSnondim}\\
\nablab\cdotb\bm{u} &=& 0,\label{eqn:boussincompnondim}\\
b &=& T-S,\label{eqn:boussstatenondim}
\end{eqnarray}
where $b(\xb,t)$ is the nondimensional buoyancy field. These equations depend on the diffusivity ratio $\tau = \kappa_T/\kappa_S$ and the Prandtl number $\sigma = \nu/\kappa_T$. The dimensionless height of the domain is $H = h/L$.

We now present a one-dimensional model for double-diffusive layering of the form~\eqref{eqn:generalg}--\eqref{eqn:generale}, developed using a horizontal averaging process. Oceanic observations of double-diffusive staircases show a horizontal extent far greater than the thickness of the individual layers \citep[e.g.][]{Schmitt_C-SALT}. As such, a horizontally averaged one-dimensional model is appropriate, providing insight to the physics of layering within a model that is relatively simple computationally. We apply the averaging process employed by \citet{Pruzina2022stirred}, detailed in appendix~\ref{sec:modelderivation}, to obtain the following system:
\begin{eqnarray}
T_t &=& \left(\frac{l^2e}{le^{1/2}+1}T_z\right)_z,\label{eqn:ddcT}\\
S_t &=& \left(\frac{l^2e}{le^{1/2}+\tau}S_z\right)_z,\label{eqn:ddcS}\\
e_t &=& \left(\frac{l^2e}{le^{1/2} + \sigma}e_z\right)_z - \sigma\left(\frac{l^2e}{le^{1/2}+1}T_z - \frac{l^2e}{le^{1/2}+\tau}S_z\right) + \sigma e_{zz} - \epsilon \frac{e^{3/2}}{l},
\label{eqn:ddce}
\end{eqnarray}
where $T$, $S$ and $e$ represent the horizontally averaged temperature, salinity and turbulent kinetic energy. The parameter $\epsilon$ controls the strength of the dissipation term. The system is closed by the mixing length $l$. The parameterisation of $l$ is critical to the model, as it controls the form of the flux terms, and therefore the nature of any instability (cf. \S\ref{sec:3Phillips}). 

It should be noted that this is a model of turbulent flow, relying on parameterisations of fluxes in terms of eddy diffusivities. As such, it should not be expected, and is not intended, to describe non-turbulent states accurately.
Numerical simulations \citep[e.g.][]{Stellmach_et_al_2011} show that the salt fingering instability quickly leads to a highly turbulent state; although this form of model cannot capture the initial salt fingering stage of the evolution, it is nonetheless appropriate to describe all subsequent development, including the formation and evolution of staircases.
Despite the fact that we are modelling a turbulent system, observations and numerical simulations show that the diffusivity ratio $\tau$ is critical to the evolution of double-diffusive staircases, and hence it is important to keep it in the model. As such, the eddy diffusivities are not identical in each equation, as may be expected in a fully turbulent model. If $\tau = 1$, there would be no difference between the two components of buoyancy, and layering would not be possible without an external forcing. 

The aim of our model is to describe layering within a mixing-length framework, through the instability mechanism described in \S\,\ref{sec:3Phillips}, which we have shown to be mathematically equivalent to the $\gamma$-instability of \citet{Radko_2003} in the case where the flux functions are parameterised in terms of the density ratio $R=T_z/S_z$. The $\gamma$-instability theory depends on the parameterisation of fluxes specifically in terms of $R$, rather than in terms of the gradients individually.  With this in mind, we propose to parameterise the mixing length $l$ in terms of $R$. From a physical perspective, the length scale can be interpreted as a characteristic size of turbulent eddies, indicating that it should depend on $e$ as well as $R$.  

For a layered system, the mixing length must be parameterised such that $l$ is small in the narrow, high-gradient interfaces, and larger in the wider, well-mixed layers. 
In salt fingering systems, it has been established, both numerically and experimentally, that the temperature and salinity fluxes have a decreasing dependence on the density ratio $R=T_z/S_z$ \citep[e.g.][]{mcdougall1984flux,kimura2011turbulence}. Furthermore, for the $\gamma$-instability to be present, the flux ratio $\gamma=f/c$ has a decreasing dependence on the density ratio $R=T_z/S_z$ in the equilibrium states. In our system (3.15)--(3.17), the temperature flux is $f = l^2eT_z/\left(le^{1/2}+1\right)$ and the salt flux is $c = l^2eS_z/\left(le^{1/2}+\tau\right)$. From these forms, $f$, $c$ and $\gamma$ are all increasing functions of the length scale $l$, and hence for $f(R)$, $c(R)$ and $\gamma(R)$ to be decreasing functions, $l(R)$ must also be decreasing.
Mathematically, for a system of the form~\eqref{eqn:generalg}--\eqref{eqn:generale}, a choice of $l$ that depends on $R$ alone leads to the high-wavenumber instability discussed in \S\,\ref{sec:3Phillips}, forming an ill-posed problem. Therefore, we parameterise $l$ also to include a dependence on the local kinetic energy $e$. Based on these considerations, we adopt the parameterisation
\begin{equation}
l = \frac{\left(e^2 + \delta R^2\right)^{1/2}}{e^{1/2}R},\label{eqn:ddclength}
\end{equation}
where $\delta$ is a  parameter  chosen to be small such that, for $O(1)$ values of $e$ and $R$, the mixing length $l\sim\sqrt{e}/{R}$. The value of $\delta$ must be non-zero, as otherwise $e=0$ is a steady-state solution for all values of $R_0$, susceptible to the energy-mode instability~\eqref{eqn:generalcond1}. For values of $R$ close to unity, the prescription~\eqref{eqn:ddclength} is a decreasing function of $R$, giving a large length when $R\approx 1$, and smaller lengths for larger values of $R$, as required. We note that, while the model \eqref{eqn:ddcT}--\eqref{eqn:ddce} was derived via averaging processes and physical arguments, by contrast there is not such a clear physical motivation for the exact form of the length scale. The prescription \eqref{eqn:ddclength} is therefore not the only possible choice, but is, nonetheless, a simple parameterisation with the appropriate qualitative dependence on $e$ and $R$ to model layers.

It should be noted that in  double-diffusive convection described by the Boussinesq equations \eqref{eqn:boussu}--\eqref{eqn:boussstate}, the buoyancy anomalies leading to convective motions appear at or below the salt fingering length scale ($l=1$, in our nondimensionalisation), while  turbulent mixing is expected to occur on scales larger than this. Our turbulent model cannot describe these small scale dynamics directly, and instead assumes that turbulent motion continues down to the smallest scales to parameterise these motions. The length measures the scale of the turbulent motion, rather than the layers themselves, so that in these strongly convective regions, we expect $l$ to represent the size of a turbulent eddy, which may be significantly smaller than the full layer depth.

At this stage, it is helpful to review the parameter values for which the two different regimes of double-diffusive convection occur. In the salt fingering regime, both temperature and salt gradients are positive. For the fluid to be statically stable, it is  required that $b_z = 1 - 1/R_0 > 0$, and hence the background density ratio
\begin{equation}
R_0 > 1.
\end{equation}
For diffusive convection, both gradients are negative. The overall buoyancy gradient is $b_z = -1 + 1/R_0$, so for the fluid to be statically stable, it is required that
\begin{equation}
0<R_0 <1.
\end{equation}

The system~\eqref{eqn:ddcT}--\eqref{eqn:ddclength} depends on four dimensionless parameters: $\tau$, $\sigma$, $\delta$ and $\epsilon$. The first two are material parameters of the fluid under consideration. For example, salt water has a diffusivity ratio of $\tau\approx0.01$ and a Prandtl number of  $\sigma=O(10)$, depending on the temperature and salinity. In this study, we focus primarily on illustrating the results of our model for the case of salt water, but we note that the choice of $\tau$ and $\sigma$ can be  adapted to model other physical contexts. The latter two parameters, $\delta$ and $\epsilon$, are empirical modelling parameters introduced in the derivation to control the relative importance of turbulent dissipation and the form of the mixing length \eqref{eqn:ddclength} (the parameter $\epsilon$ also appears in the model of BLY). We will determine values of $\delta$ and $\epsilon$   that lead to physically realistic behaviour in the solutions.

\begin{figure}
\centerline{\includegraphics[width=\textwidth]{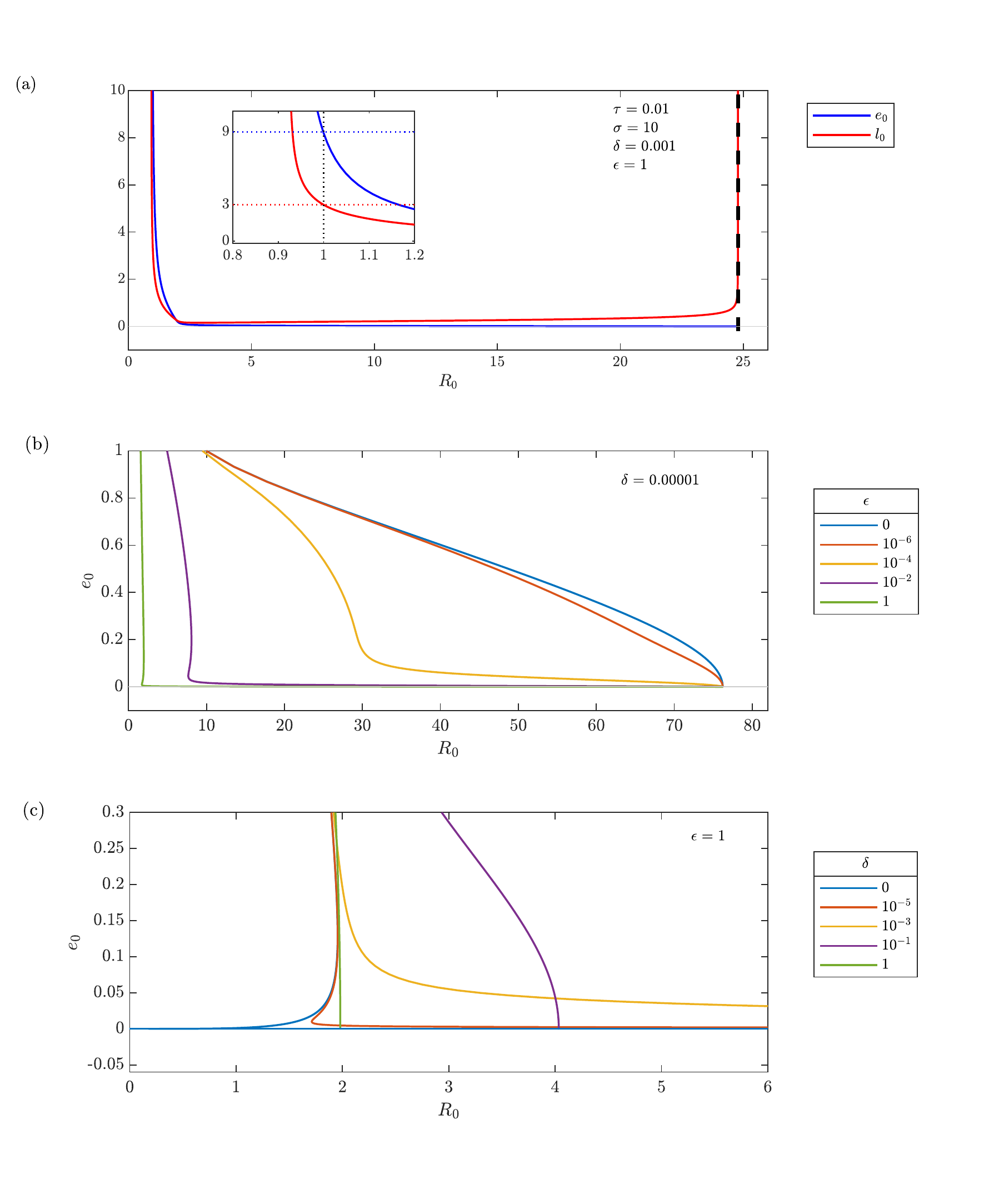}}
\caption{Steady-state solutions to~\eqref{eqn:steadye} with $\tau = 0.01$, $\sigma = 10$. ($a$) Energy $e_0$ and  corresponding value of the length scale $l_0$ given by~\eqref{eqn:ddclength}, as functions of $R_0$ for $\epsilon = 1$, $\delta = 0.001$. For small $R_0$, $e_0$ and $l_0$ are large; as $R_0$ increases, $e_0$ decreases, with $l_0$ initially decreasing but $l_0\to\infty$ as $e_0\to0$. Inset plot shows behaviour of $e_0$ and $l_0$ near $R_0 = 1$, with red and blue dotted lines showing the values $e_0 = \sigma/\epsilon-1 = 9$ and $l_0 = \sqrt{\sigma/\epsilon-1} = 3$. ($b$) $e_0$ as a function of $R_0$, for a range of values of $\delta$ with  $\epsilon = 1$ fixed. ($c$) $e_0$ as a function of $R_0$, for a range of values of $\epsilon$ with $\delta = 0.001$ fixed. Sufficiently small values of $\delta$ and large values of $\epsilon$ lead to $e_0(R_0)$ being multi-valued.}
\label{fig:ddcsteadyenergies}
\end{figure}

\section{Steady states and their stability}\label{sec:stability}
To analyse the stability of the system~\eqref{eqn:ddcT}--\eqref{eqn:ddce}, we begin by applying the general linear stability theory of three-component systems developed in \S\,\ref{sec:3Phillips}. Equations~\eqref{eqn:ddcT}--\eqref{eqn:ddce} may be expressed in the form~\eqref{eqn:generalg}--\eqref{eqn:generale} by writing
\begin{eqnarray}
f &=& \frac{l^2e}{le^{1/2} + 1}g,\label{eqn:specificf}\\
c &=& \frac{l^2e}{le^{1/2} + \tau}d,\label{eqn:specificc}\\
\kappa &=& \frac{l^2e}{le^{1/2} + \sigma} + \sigma,\\
 p &=&  - \sigma\left(\frac{l^2e}{le^{1/2}+1}g - \frac{l^2e}{le^{1/2}+\tau}d\right) - \epsilon \frac{e^{3/2}}{l},\label{eqn:specificp}
\end{eqnarray}
where $g = T_z$ and $d = S_z$. For a given value of $R_0$, the system admits the uniform steady state $(g_0,d_0,e_0) = (\pm1,\pm1/R_0,e_0(R_0))$.

\subsection{Steady states}

With the length scale prescribed by~\eqref{eqn:ddclength}, the steady-state equation $p=0$ for $e_0(R_0)$ leads to the following algebraic relation between $e_0$ and the parameter $R_0$:
\begin{multline}
	g_0\left(R_0-1\right)\left(e_0^2+\delta R_0^2\right)^2 + g_0\left(\tau R_0-1\right)\left(e_0^2+\delta R_0^2\right)^{3/2}R_0\\ + \frac{\epsilon}{\sigma}R_0^3e_0^2\left(e_0^2+\delta R_0^2\right) + \left(1+\tau\right)\frac{\epsilon}{\sigma}\left(e_0^2+\delta R_0^2\right)^{1/2}e_0^2  + \frac{\epsilon}{\sigma}R_0^5\tau e_0^2 = 0,
\label{eqn:steadye}
\end{multline}
where $g_0 = \pm 1$. We interpret the uniform-gradient steady state physically as a representation of the flow resulting from salt fingers. Individual fingers cannot be distinguished, but a mean fluid motion is supported by one stable and one unstable component of the buoyancy gradient.

Figure~\ref{fig:ddcsteadyenergies}($a$) shows $e_0$ as a function of $R_0$. When there is no overall stratification ($R_0=1$), the energy $e_0\approx \sigma/\epsilon -1$ (shown by the blue dotted line in the inset plot).  As $R_0$ increases, $e_0$ decreases, reaching $e_0=0$ at $R_0 = (1+\sqrt{\delta})/(\tau + \sqrt{\delta})$. However, it is important to note that such an unphysical `zero energy turbulent state' is precluded in our time-dependent model.
To see this, we consider  the evolution of a state that begins with  $e>0$ for all $z$. At a given time, let $z = z_*$ denote the  position of a local minimum of $e$, with $e_z(z_*)=0$, $e_{zz}(z_*)>0$, and $e(z_*)$ is near zero. Additionally, we note from~\eqref{eqn:ddclength} that $le^{1/2} \to \sqrt{\delta}$ as $e\to0$. After some algebra, the governing energy equation \eqref{eqn:ddce} at $z = z_*$ gives
\begin{equation}
e_t(z_*) \sim \left(\frac{\delta}{\sqrt{\delta}+\sigma} + \sigma\right)e_{zz} - \sigma\left(\frac{\delta T_z}{\sqrt{\delta}+1} - \frac{\delta S_z}{\sqrt{\delta}+\tau}\right).
\label{eq:et}
\end{equation}
Provided that $R_0<(\sqrt{\delta}+1)/(\sqrt{\delta}+\tau)$ (i.e.\ $R_0$ is in the range where uniform steady states exist), every term on the right-hand side of \eqref{eq:et} is positive. It follows that $e_t(z_*)>0$, implying that the minimum energy cannot decrease, precluding the energy from ever reaching zero. Hence,  while zero energy and negative energy states can exist within the equations, they will never be attained from initial conditions starting  with positive energy. Nonetheless, the change of sign of $e_0$ when $R_0>(\sqrt{\delta}+1)/(\sqrt{\delta}+\tau)$ means that this model is applicable only for sufficiently low density ratios.

Figure~\ref{fig:ddcsteadyenergies}(a) also shows the value of the length scale $l_0$ calculated from the energy using~\eqref{eqn:ddclength}. As $R_0\to 1$, $l_0\sim \sqrt{e_0}\approx \sqrt{\sigma/\epsilon - 1}$, which is shown by the red dotted line in the inset. As $R_0$ increases from $1$, the length scale initially decreases, reaching a minimum, before increasing, with $l_0\to\infty$ and $e_0\to0$ as $R_0\to(\sqrt{\delta}+1)/(\sqrt{\delta}+\tau)$. However, we have shown that the zero energy state is never reached, and hence this divergence in the mixing length will never occur. 

The relationship between $e_0$ and $R_0$ given by \eqref{eqn:steadye} is shown for various choices of the  parameters $\epsilon$ and $\delta$ in figures~\ref{fig:ddcsteadyenergies}($b$)--($c$). The results illustrate a strong dependence of $e_0$ on both $\epsilon$ and $\delta$, with smaller values of $\delta$ and larger values of $\epsilon$ causing $e_0$ to be multi-valued for some range of $R_0$. For example, the solution shown in purple in figure~\ref{fig:ddcsteadyenergies}($b$) is multi-valued near $R_0 \approx 8$. In cases where there are multiple steady-state energies, one steady state is unstable to the energy mode, leading to growth in energy on the domain scale. To investigate layering processes we wish to avoid this situation, so we set $\epsilon = 1$ and $\delta = 0.001$. These are the parameters used in figure~\ref{fig:ddcsteadyenergies}($a$), where $e_0$ is clearly single-valued throughout.

In the diffusive convection regime, $T_z, S_z>0$ and $0<R_0<1$. Within these ranges, all the terms in~\eqref{eqn:steadye} are negative, and hence there are no positive solutions for $e_0$ in the diffusive convection regime. This result holds in general for any system of the form~\eqref{eqn:specificf}--\eqref{eqn:specificp}, no matter what parameterisations are adopted for the length scale and dissipation term. As such, it appears that an unforced system of this form is not sufficient to model layering in diffusive convection, lending weight to the proposition of \citet{ma2022thermohaline} that external forcing may be necessary. In this paper, we will restrict our focus to the salt fingering regime, with the modelling of layering in diffusive convection providing an interesting problem for future work.

\subsection{Linear stability}
\begin{figure}
\centerline{\includegraphics[width=0.88\textwidth]{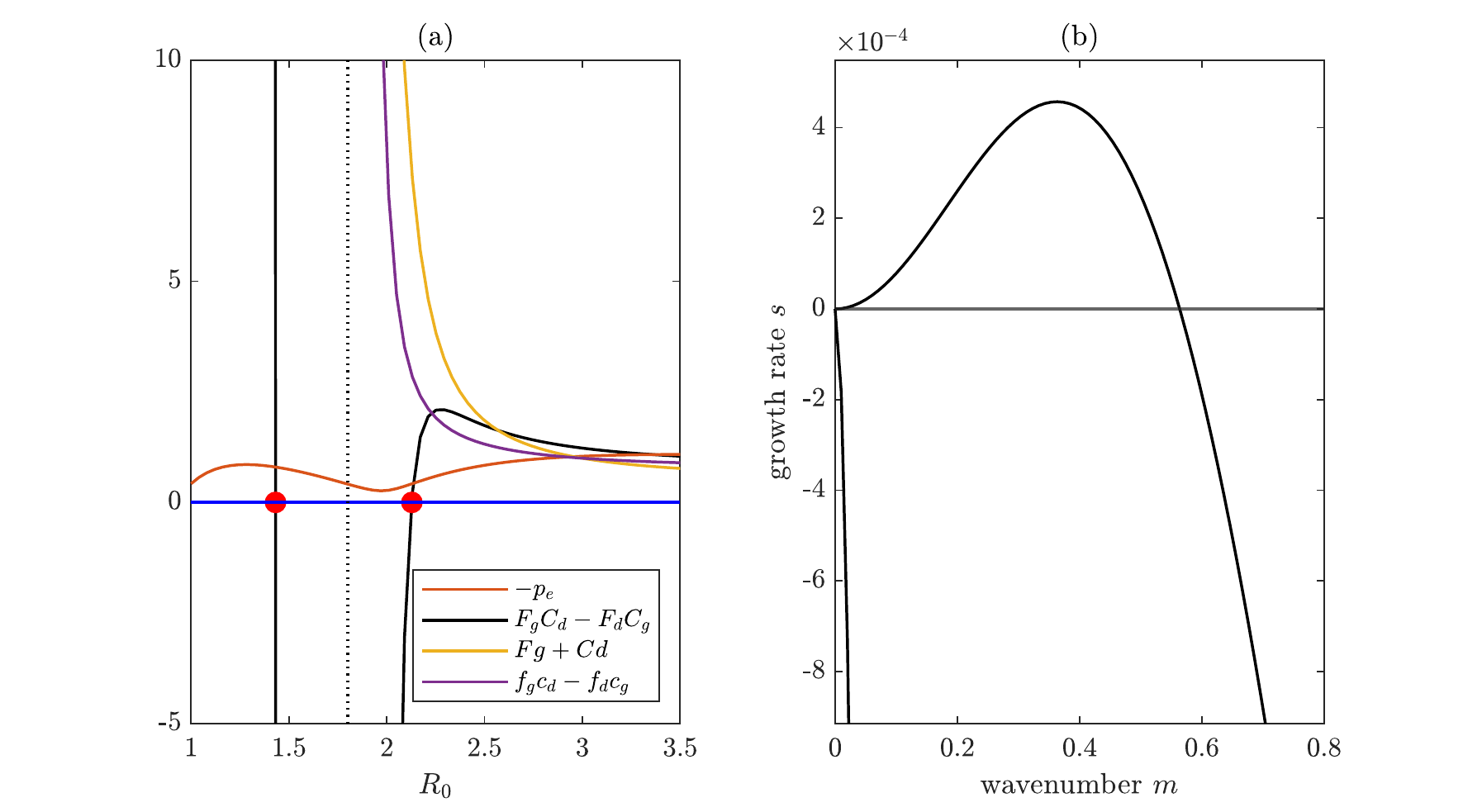}}
\caption{($a$) Various stability measures of the uniform steady state as $R_0$ varies, with $\tau = 0.01$, $\sigma = 10$, $\epsilon = 1$, $\delta = 0.001$. The quantity $-p_e$ is shown in red,  $F_gC_d - F_dC_g$ in black, $F_g + C_d$ in yellow, and $f_gc_d - f_dc_g$ in purple. The red circles mark the minimum and maximum values of $R_0$ for which condition \eqref{eqn:generalcond2} is satisfied. ($b$) The growth rate $s$ against wavenumber $m$ for the steady state with $R_0=1.8$ (marked with a dotted black line in ($a$)). There is a single unstable mode with maximum growth rate $s=4.6\times10^{-4}$ at wavenumber $m=0.363$.}
\label{fig:ddcinstability}
\end{figure}
The stability of the steady states of~\eqref{eqn:ddcT}--\eqref{eqn:ddce} can be analysed using the framework described in \S\,\ref{sec:3Phillips}. For a range of values of $R_0$, we first calculate $e_0(R_0)$ and substitute the value into the expressions for $-p_e$, $F_gC_d - F_dC_g$, $F_g + C_d$ and $f_gc_d-f_dc_g$; these quantities are plotted as functions of $R_0$ in figure~\ref{fig:ddcinstability}($a$). For a finite range of $R_0$ (between the red dots), $F_gC_d-F_dC_g<0$, thereby satisfying the condition for the Phillips instability. By comparison with the schematic in figure~\ref{fig:3eqnstab}($b$), we see that our expectation of a finite unstable region is met.  Note that by our choice of nondimensionalisation, $|\Ra| = 1$, so varying $R_0$ in figure~\ref{fig:ddcinstability}($a$) represents a single path through the unstable region of $\Ra$--$\Rs$ space. The values of $\epsilon$ and $\delta$ are chosen so that the energy mode is not unstable (condition~\eqref{eqn:generalcond1}), and neither condition~\eqrefl{eqn:generalcond2}{b} nor \eqref{eqn:generalcond4} is met.
Figure~\ref{fig:ddcinstability}($b$) shows a plot of growth rate against wavenumber for the case of $R_0 = 1.8$ --- there is a single unstable mode, with a uniquely defined wavenumber of maximum growth rate, which can be used to predict the width of the fastest growing perturbations, and hence the width of the initial layers formed.

Recall from the end of \S2 that in a system of the general form~\eqref{eqn:generalg}--\eqref{eqn:generale}, the layering instability (condition \eqrefl{eqn:generalcond2}{a}) may be caused by either the forcing mechanism of BLY and \cite{PvH}, or the $\gamma$-instabiliy of \citet{Radko_2003}. With the specific system~\eqref{eqn:ddcT}--\eqref{eqn:ddce}, this is no longer the case. For a model with no source term in the energy equation~\eqref{eqn:ddce}, the $\gamma$-mechanism is the only one in play.

\begin{figure}
    \centerline{\includegraphics[width=0.99\textwidth]{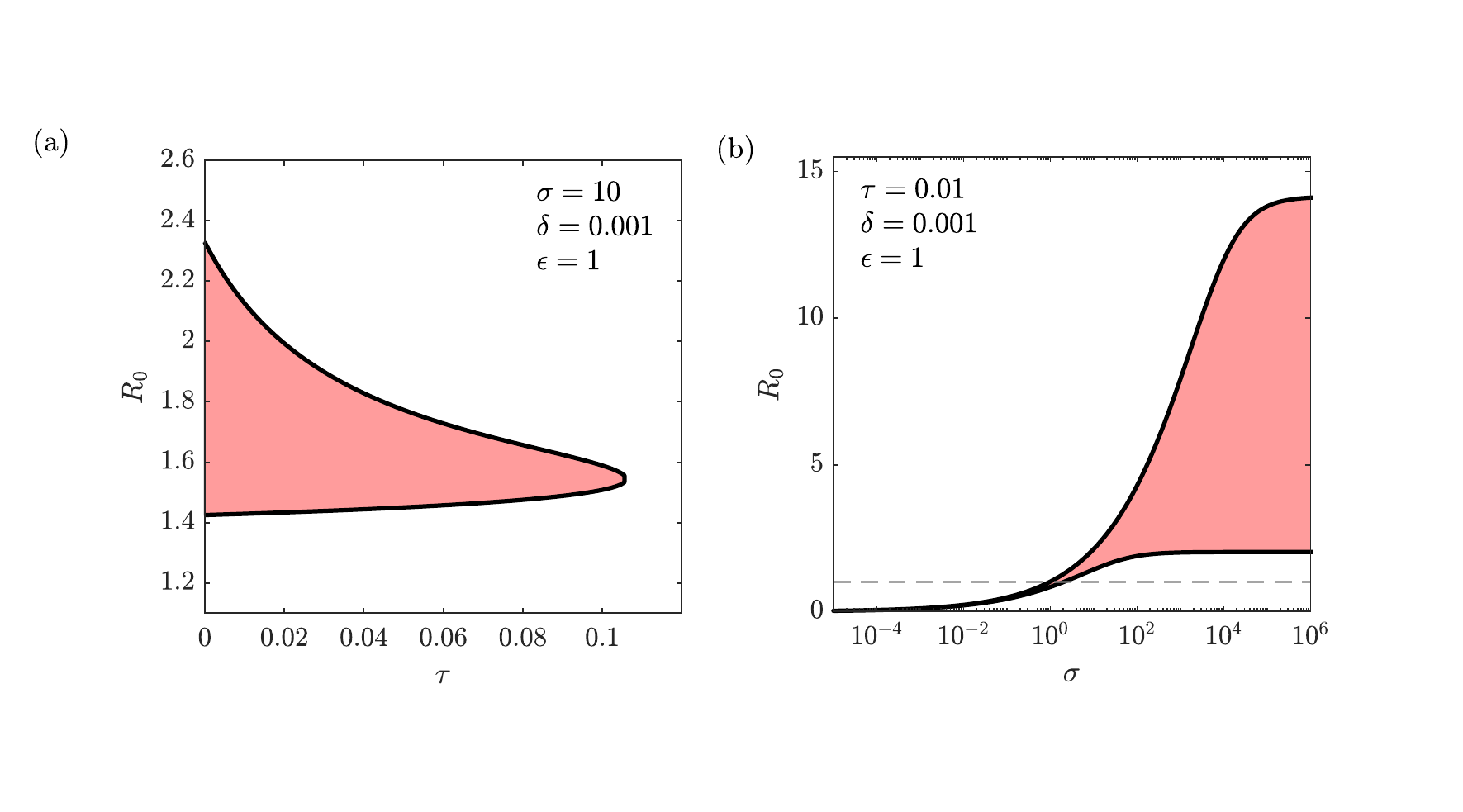}}
    \caption{Effect on the layering instability of changing (a) the diffusivity ratio $\tau$, and (b) the Prandtl number $\sigma$. Black lines show the boundary of the unstable range of $R_0$, as ($a$) $\tau$ is changed with fixed $\sigma = 10$, and ($b$) $\sigma$ is changed for fixed $\tau = 0.01$. In each case, changing the value of the other parameter leads to no qualitative differences. The other model parameter values are $\delta = 0.001$ and $\epsilon = 1$; for these choices of $\delta$ and $\epsilon$, there is no energy mode instability. The dashed line in ($b$) shows $R_0=1$, the lower boundary of the salt fingering regime.}
    \label{fig:changetausigma}
\end{figure}

We now investigate the effect on the stability of the system of varying the values of the material parameters, while fixing $\delta = 0.001$ and $\epsilon = 1$. Figure~\ref{fig:changetausigma} shows the effect on the critical values of $R_0$ for instability of changing $\tau$ and $\sigma$ independently. The black line in figure~\ref{fig:changetausigma}($a$) shows the minimum and maximum values of $R_0$ for which instability occurs, as $\tau$ is varied with all other parameters kept fixed. The critical value of $\tau$ at the tip of the curve is $\tau_c = 0.1055$. This critical value is independent of $\sigma$, although larger values of $\sigma$ lead to larger unstable ranges of $R_0$. Figure~\ref{fig:changetausigma}($b$) shows the effect of varying $\sigma$ on the critical values of $R_0$.
For $\sigma\ll1$, only a very narrow range of $R_0$ leads to instability; at larger values of $R_0$ this range increases significantly, saturating at approximately $2\lesssim R_0\lesssim 14$ for large $\sigma$. The dashed line shows the lower boundary of the  fingering regime, at $R_0=1$. Larger values of $\tau$ reduce the size of the unstable range of $R_0$, but there is little qualitative change. For small $\sigma$, the entire unstable range lies below $R_0=1$, and is therefore not in the salt fingering regime. This result is consistent with those of  \citet{traxler2011numerically}, who studied salt fingering at low Prandtl number using three-dimensional  numerical simulations. \citet{traxler2011numerically} found that the empirical flux ratio $\gamma$ increased monotonically with density ratio $R$, so layering by the $\gamma$-instability was not expected at small $\sigma$. Instead, it was suggested that any layering was due to the collective instability of \citet{stern_1969}.

\section{Nonlinear evolution and long-term merger behaviour}\label{sec:numerics}
A key advantage of the regularisation inherent in our mixing-length formulation  \eqref{eqn:ddcT}--\eqref{eqn:ddce}  is that it allows the investigation of long-term dynamics of staircase evolution and merger beyond any initial instability. To investigate this long-term behaviour, we solve the system~\eqref{eqn:ddcT}--\eqref{eqn:ddce} with length scale~\eqref{eqn:ddclength} subject to the boundary conditions 
\begin{eqnarray}
T(0) &=& 0,\qquad T(H) =  H,\label{eqn:bcT}\\
S(0) &=& 0,\qquad S(H) = H/R_0,\label{eqn:bcS}\\
e(0) &=& e_0,\qquad e(H) = e_0,\label{eqn:bce}
\end{eqnarray}
and initial conditions
\begin{eqnarray}
T &=&  z - g'\sin\left(\frac{2n\pi z}{H}\right),\label{eqn:icT}\\
S &=&  \frac{z}{R_0} - d'\sin\left(\frac{2n\pi z}{H}\right),\label{eqn:icS}\\
e &=& e_0 - e'\frac{2n\pi}{H}\cos\left(\frac{2n\pi z}{H}\right),
\label{eqn:ice}
\end{eqnarray}
where the perturbation amplitudes $(g',d',e')$ are chosen such that the initial condition is an eigenstate of the linear stability problem, with $2n\pi/H$ chosen to be the maximally unstable wavenumber based on the linear theory of \S~\ref{sec:stability}. We solve the model numerically using the MATLAB \texttt{pdepe} solver, with $4000$ spatial mesh points across a domain of depth $500$, which is sufficient for well-resolved solutions. The solver varies time steps dynamically to ensure adequate resolution. 

\subsection{Nonlinear evolution}
\begin{figure}
\centering
\centerline{\includegraphics[width=0.999\textwidth]{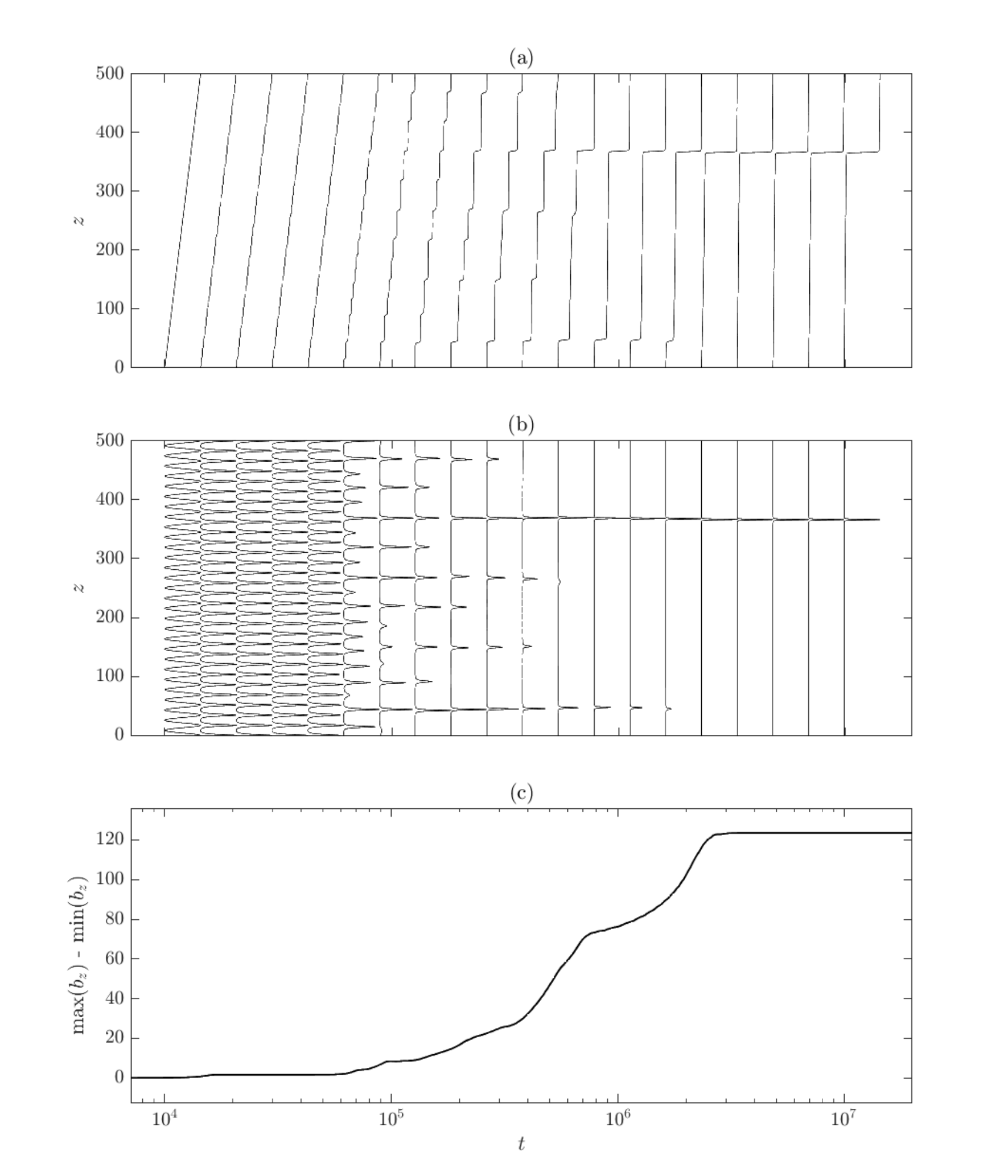}}
\caption{Nonlinear evolution of the system~\eqref{eqn:ddcT}--\eqref{eqn:ddce}, with length scale~\eqref{eqn:ddclength}, subject to boundary conditions~\eqref{eqn:bcT}--\eqref{eqn:bce} and initial conditions~\eqref{eqn:icT}--\eqref{eqn:ice}, for parameter values $\tau = 0.01$, $\sigma = 10$, $\delta = 0.001$, $\epsilon = 1$, $R_0 = 1.8$. ($a$) Depth profiles of the overall buoyancy field $b=T-S$ at a range of times logarithmically distributed between $t=10^4$ and $t=10^7$; ($b$) profiles of the buoyancy gradient $b_z = T_z-S_z$ scaled by the range of its values at each time, plotted at the same times as in ($a$); ($c$) range of gradients, i.e\ $\text{max}(b_z)-\text{min}(b_z)$. The solution evolves from the initial condition into a dense stack of layers (seen as the first solution presented in ($b$)). At $t\approx 6\times10^5$, the layers begin to undergo mergers, which cause the maximum gradient to increase, until by $t\approx 2\times10^6$, only a single interface remains at $z\approx 350$, with $b_z$ $\approx 120$.}
\label{fig:ddcresults}
\end{figure}

 Figure~\ref{fig:ddcresults} shows the results of a numerical solution of \eqref{eqn:ddcT}--\eqref{eqn:ddce}, for the same parameters as used in figure~\ref{fig:ddcinstability}($b$). Figure~\ref{fig:ddcresults}($a$) shows the buoyancy field $b=T-S$ plotted over a range of times.  The plot illustrates the evolution from an initially uniform gradient to a layered staircase; the layers then proceed to undergo mergers over time. Eventually, at $t\approx 2\times10^6$, only a single interface remains, located at $z\approx 350$. Figure~\ref{fig:ddcresults}($b$) shows the normalised buoyancy gradient at the same time points. Interfaces between layers are represented by sharp spikes in the gradient profile; these profiles reveal the fine structure at early times that is not visible in the overall temperature field in figure~\ref{fig:ddcresults}($a$). The range in the buoyancy gradient, i.e.\ $\text{max}(b_z)-\text{min}(b_z)$, is shown in figure~\ref{fig:ddcresults}($c$), measuring the difference between the gradients in the interfaces and the layers. There is a clear gradual increase in the maximum gradient, beginning at $t\approx6\times10^5$ and ending at $t \approx 2 \times 10^6$, defining the range of times over which mergers occur. During a layer merger, the overall buoyancy variation across a region must be conserved, resulting in sharper interfaces with higher gradients after each merger. Referring to the unstable region shown in figure~\ref{fig:ddcinstability}($a$), there is no constraint individually on $T_z$ and $S_z$, provided that $R_0$ stays within the bounds of the unstable region. Hence $b_z$ can become arbitrarily large, resulting in ever steeper interfaces as successive merger events take place. This contrasts with results obtained for related models of stirred, one-component layering  \citep{BLY98,Pruzina2022stirred}, where a well-defined maximum unstable gradient is determined, such that subsequent mergers produce thicker interfaces of fixed gradient. The long-term evolution of merger behaviour follows the same inverse logarithmic trend identified by \cite{Pruzina2022stirred}, with the number of remaining interfaces $N$ obeying the scaling $1/N\sim \log(t)$ as $t \to \infty$.

\subsection{Merger dynamics}

To understand the merging behaviour in more detail, we consider the results in the context of the analysis of  \citet{radko2007mechanics}, who identified two types of mergers: the B-merger, in which relatively strong interfaces grow at the expense of weaker neighbouring interfaces, and the H-merger, where neighbouring interfaces drift and collide. By considering a one-dimensional buoyancy conservation equation in a stepped basic state, and analysing the variation of the buoyancy flux across a step, Radko demonstrated the so-called merger theorem, showing that the B-merger has growth rate $\lambda_B\propto -\tpd{\tilde{F}}{\Bt}$, and the H-merger has growth rate $\lambda_H\propto \tpd{\tilde{F}}{\Ht}$, where $\tilde{F}(\Bt,\Ht)$ is the buoyancy flux across a step, $\Bt$ the buoyancy jump, and $\Ht$ the height of the step.

To apply the merger theorem to the BLY model, \citet{radko2007mechanics} considered constant flux solutions, for which the model can be reduced to a  nonlinear oscillator equation for $e(b)$. To apply the same analysis to our model, we adopt the same approach. Thus we seek steady-state solutions to the system~\eqref{eqn:ddcT}--\eqref{eqn:ddce} with uniform temperature and salinity fluxes $f_0$ and $c_0$. The system can then be reduced to a single equation (see Appendix~\ref{sec:oscillatorderivation}), given by
\begin{equation}
\tsfrac{1}{2}e_T^2 + U(D) = E\left(\frac{ D^2}{\kappa f_0 \left(D+1\right)}\right)^2,\label{eqn:nussoscillator}
\end{equation}
where $D(e) = le^{1/2}$, the potential $U(D)$ is defined by
\begin{equation}
U(D) = \left(\frac{D^2}{\kappa f_0 \left(D+1\right)}\right)^2 \bigintsss \left( -\sigma\left(f_0-c_0\right) - \epsilon\frac{e(D)^2}{D} \right)\kappa \intd  D,\label{eqn:nusspotential}
\end{equation}
and  $e(D)$ is defined by
\begin{equation}
e(D) = \frac{\sqrt{D^2-\delta}\left(D+1\right)}{D+\tau}\gamma_0.\label{eqn:nusse(D)}
\end{equation}
 Equation~\eqref{eqn:nussoscillator} represents a nonlinear oscillator for $e$ as a function of temperature $T$, with variable weight. By inverting~\eqref{eqn:nusse(D)} for $D(e)$, \eqref{eqn:nussoscillator} is transformed into an equation for $e_T$ in terms of $e$. Analytically, this requires writing $D(e)$ as the root of a quartic, but the inversion is simple to do numerically by coupling~\eqref{eqn:nusse(D)} with~\eqref{eqn:nusspotential}. 

\begin{figure}
\centerline{\includegraphics[width=0.99\textwidth]{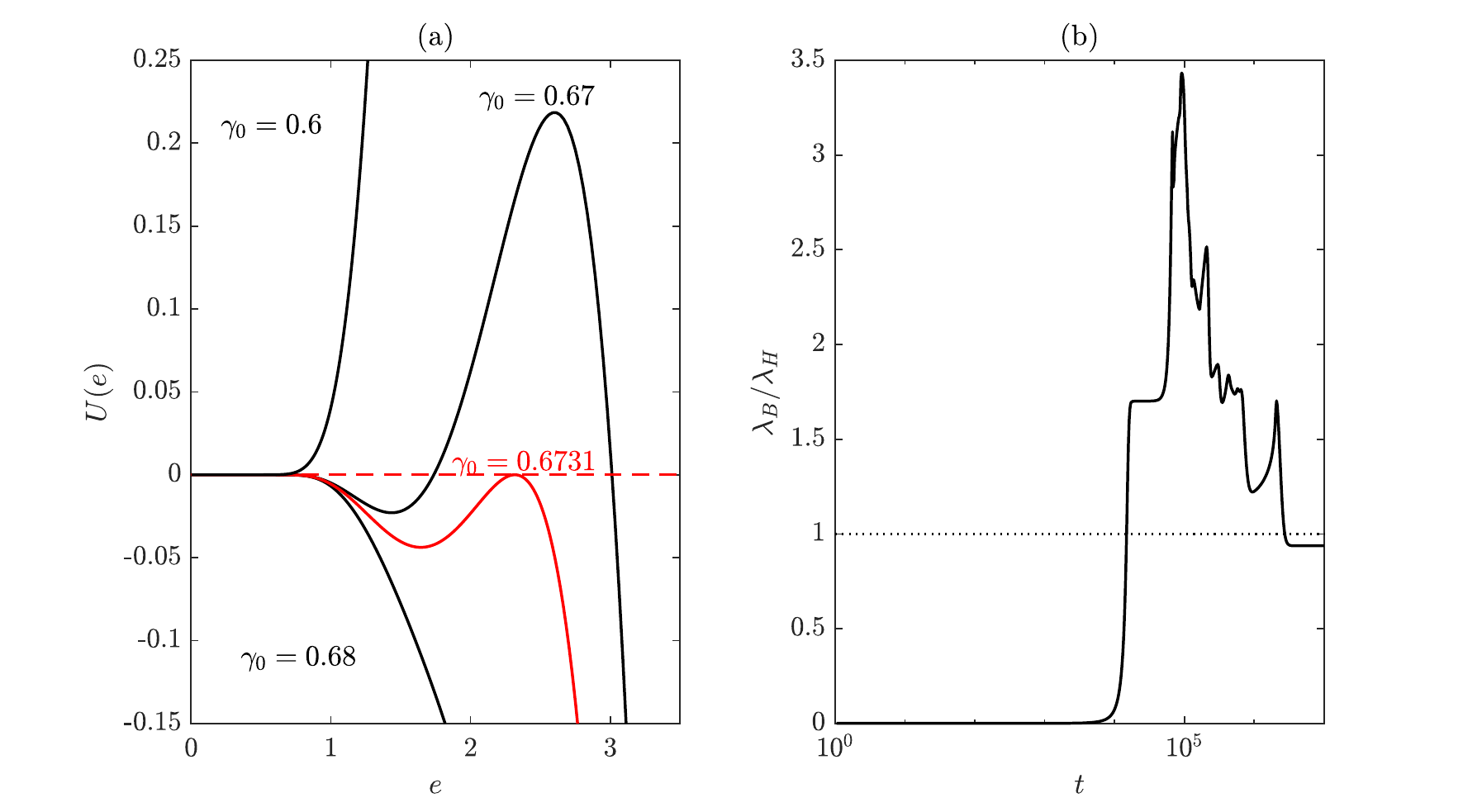}}
\caption{($a$) Potential $U(e)$ found by integrating~\eqref{eqn:nusspotential} with respect to  $D$, and coupling with \eqref{eqn:nusse(D)} to find corresponding values of $e$. The red line has two peaks at $e_1$ and $e_2$, where $U(e_1) = U(e_2)$. Parameter values are $f_0 = 0.45$, $\tau = 0.01$, $\sigma = 10$, $\epsilon = 1$ and $\delta = 0.001$.
($b$) Plot of the ratio $\lambda_B/\lambda_H$ calculated in the small-gradient case given by equation~\eqref{eqn:grratio}, for the solutions presented in figure~\ref{fig:ddcresults}.  The dotted line marks $\lambda_B/\lambda_H = 1$: above this line, the B-merger dominates; below it, the H-merger dominates.}
\label{fig:nusspotentials}
\end{figure}

The potential $U(e)$ is plotted in figure~\ref{fig:nusspotentials}($a$). For a narrow range of values of $f_0$ and $\gamma_0$, $U(e)$ has two peaks, at $e_1$ and $e_2$. With this two-peak shape for $U(e)$, the oscillator equation~\eqref{eqn:nussoscillator} has two stable steady states corresponding to the peaks of the potential: one with a smaller energy and the other with larger energy. These correspond to values of the energy in interfaces and layers respectively. The profile of $U(e)$ depends sensitively on $\gamma_0$ and $f_0$, but for each $f_0$ there is a precise value of $\gamma_0$ such that $U(e_1) = U(e_2)$; this value of $U$ is shown by the red dashed line in figure~\ref{fig:nusspotentials}($a$). For this critical value of $\gamma_0$, the oscillator $e(T)$ has a special \textit{kink} solution linking the two maxima \citep{BLY98}. By analogy with similar kink solutions to the Cahn-Hilliard equation, more complex solutions with gradient spikes also be constructed \citep[e.g.][]{fraerman1997nonlinear}.

 With the potential taking this two-peak form, our three-component system maps directly to \citeauthor{radko2007mechanics}'s \citeyearpar{radko2007mechanics} analysis of the merging behaviour of the BLY model, which is susceptible to both B- and H-mergers. Radko shows that, in this situation, the ratio of their growth rates is
\begin{equation}
\frac{\lambda_B}{\lambda_H} = \frac{\bar{g}-g_\text{min}}{g_\text{min}},\label{eqn:grratio}
\end{equation}
where $g_\text{min}$ is the minimum gradient (in layers) and $\bar{g}$ the background gradient (averaged across the whole layer-interface system). If the ratio $\lambda_B/\lambda_H$ is greater than unity, B-mergers occur, and if the ratio is less than unity then H-mergers will dominate instead.  In the solutions shown in figure~\ref{fig:ddcresults}, it appears that the mergers occur via the B-merger pattern, with weaker interfaces shrinking without significant drifting. To show consistency with this condition on $\lambda_B/\lambda_H$, we calculate $\lambda_B/\lambda_H$ at each time using~\eqref{eqn:grratio}. We take  $g_\text{min}$ to be the global minimum gradient at each time, and $\bar{g} = 1$ as the background gradient. The ratio $\lambda_B/\lambda_H$ is shown in figure~\ref{fig:nusspotentials}($b$), where we see that for times $10^4\lesssim t\lesssim 2\times10^6$, the ratio $\lambda_B/\lambda_H>1$, implying consistency with the numerical results, in which B-mergers dominate.

\subsection{Increase of the buoyancy flux}

\begin{figure}
\centerline{\includegraphics[width=0.85\textwidth]{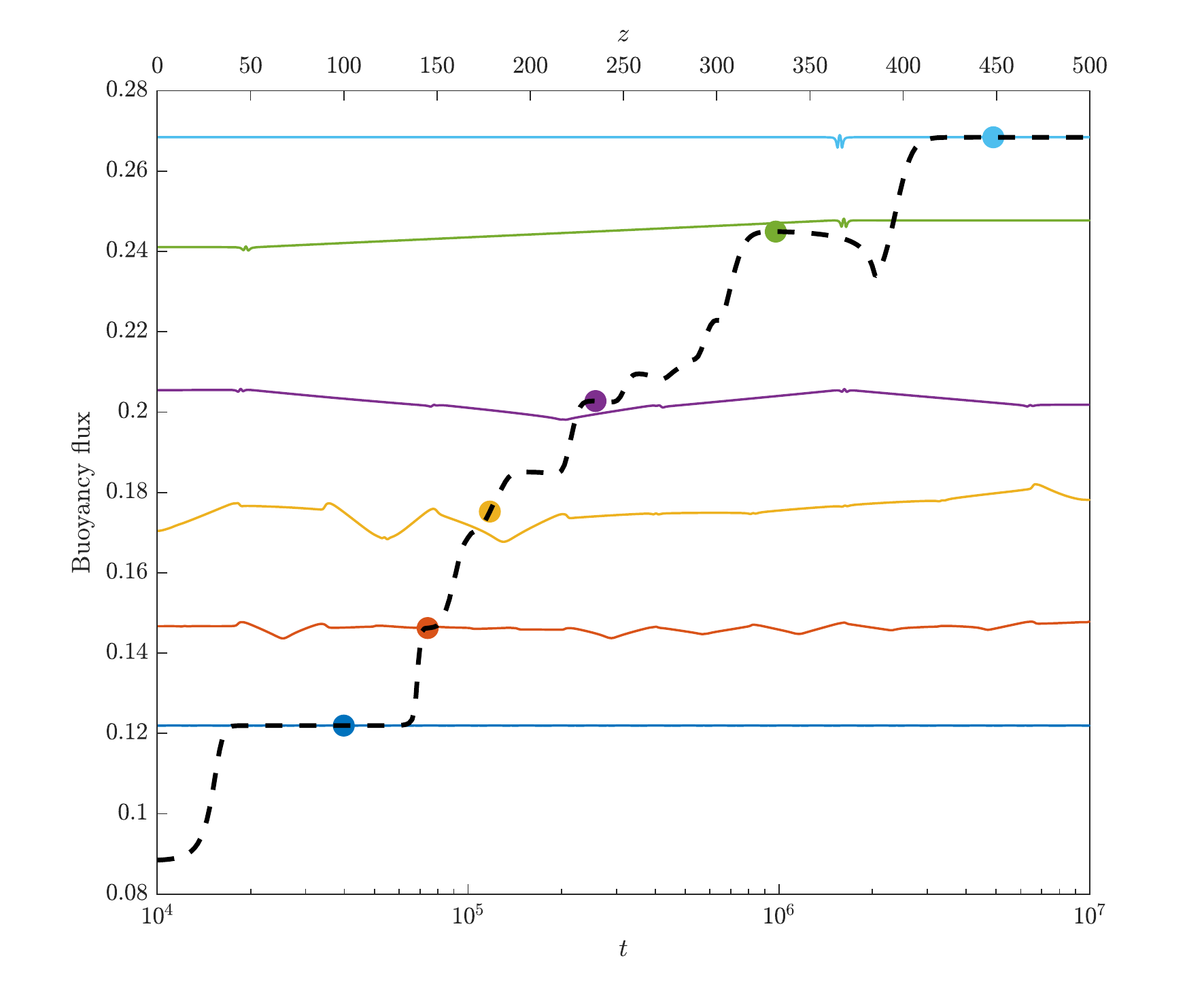}}
\caption{Evolution of the  buoyancy flux field  for the solution shown in figure~\ref{fig:ddcresults}($a$). The black dashed line shows the vertical mean of the upward buoyancy flux $(c-f)$ defined by \eqref{eq:meanflux}, plotted against time on the lower horizontal axis.  The spatial profile of the buoyancy flux (with $z$ on the upper horizontal axis) is also shown at a range of times, with the corresponding mean flux at each time marked with a dot of the same colour.}
\label{fig:buoyancyfluxtimes}
\end{figure}

In the full thermohaline system, the existence of staircases leads to greater turbulent transport of both heat and salt through the fluid in comparison with an unlayered state \citep[e.g.][]{Rosenblum_et_al_2011, HB_2021}. To investigate this effect using our model,  figure~\ref{fig:buoyancyfluxtimes} shows the evolution of the mean of the upward buoyancy flux
\begin{equation}
    \frac{1}{H} \int_0^H (c-f) \intd z,\label{eq:meanflux}
\end{equation}
calculated at each time for the solution previously shown in figure~\ref{fig:ddcresults}. There is little change in the flux at early times, until the initial linear perturbation grows into a stack of layers at $t\approx 10^4$. As layers undergo mergers, the magnitude of the flux increases, with thicker mixed layers producing larger fluxes, consistent with  simulations of salt fingering \citep[e.g.][]{Rosenblum_et_al_2011} Also plotted are the spatial profiles of the upward buoyancy flux $(c-f)$ at a range of times, showing that, in general, the flux is reasonably consistent across the different layers at each time, but with small perturbations in the interfaces.

The increase in flux seen in figure~\ref{fig:buoyancyfluxtimes} can be explained using the merger theorem of \cite{radko2007mechanics}. We saw in figure~\ref{fig:ddcresults} that layers undergo B-mergers, where weak interfaces decay with little vertical drift. To explain this, we consider a region of fluid initially containing two layers and one such weak interface. Initially, there is a  buoyancy jump  of $B_1$ across the interface. When a B-merger takes place, the resultant state is a single mixed layer, with the buoyancy variation from the top to the bottom now $B_2\ll B_1$. The merger theorem states that the system is unstable to B-mergers if the buoyancy flux decreases as the buoyancy variation increases, and so this decrease of $B$ during the merger must increase the flux in the region.

\subsection{Variation of parameter values}
\begin{figure}
\centerline{\includegraphics[width=0.999\textwidth]{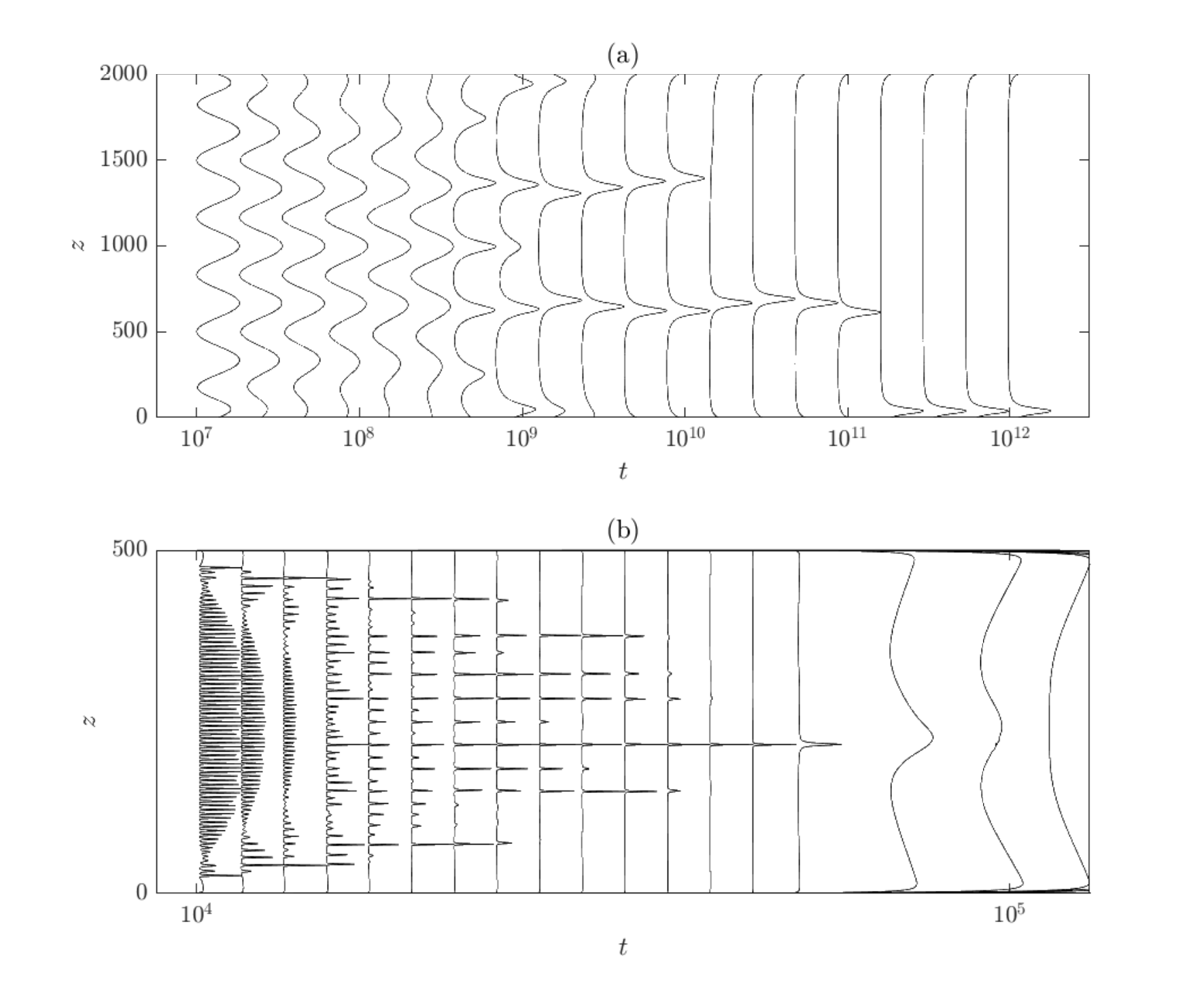}}
\caption{Evolution of solutions to~\eqref{eqn:ddcT}--\eqref{eqn:ddce} with length scale~\eqref{eqn:ddclength}, subject to boundary conditions~\eqref{eqn:bcT}--\eqref{eqn:bce} and initial conditions~\eqref{eqn:icT}--\eqref{eqn:ice}. ($a$) Very high Prandtl number case, with $\tau=0.0001$, $\sigma = 10^4$, $\delta = 0.001$, $\epsilon = 1$ and $R_0 = 3.64$, showing wide, smooth interfaces and layers. ($b$) No overall background buoyancy gradient, with $\tau = 0.01$, $\sigma = 1$, $\delta = 0.01$, $\epsilon = 1$, $R_0 = 1$, in which the timescale for mergers is similar to that for layer growth. Layers merge quickly, eventually giving way to a single convective state across the entire domain.}
\label{fig:extraresults}
\end{figure}

An exploration of parameter space reveals that
the solutions shown in figure~\ref{fig:ddcresults} are representative of a large range of parameter values. To demonstrate this, we illustrate the solution for some extreme choices of parameters, shown in figure~\ref{fig:extraresults}. Figure~\ref{fig:extraresults}($a$) takes a very large value of the Prandtl number $\sigma$, for which the scale of the most unstable mode is very large, resulting in layers and interfaces that are wider and smoother in comparison with the results of figure \ref{fig:ddcresults}. Figure~\ref{fig:extraresults}($b$) shows the case with $R_0 = 1$, on the boundary between the salt fingering and statically unstable regimes. In this case, layers do form, but mergers occur on such a fast timescale that a regular staircase never develops, and the system transitions very quickly from the linear growth phase to a single interface in the middle, which then decays, forming a single layer across the entire domain depth. In both of the cases shown, while there are some quantitative differences from the results in figure~\ref{fig:ddcresults}, the qualitative behaviour is the same.

We selected the parameter values $\delta$ = 0.001 and $\epsilon = 1$ to prevent the energy mode from being unstable. If instead we choose parameters such that there is an energy mode instability, then there is energy growth on the domain scale, producing a wide interior region with a large energy and constant temperature and salt gradients, with narrow boundary regions on either side to satisfy the boundary conditions~\eqref{eqn:bcT}--\eqref{eqn:bce}.
 
\section{Discussion}
\label{sec:conc}
We have demonstrated the development of models for staircases in three-component systems and formulated the first regularised mixing-length theory of salt fingering staircases that includes  double-diffusive physics directly.  We have shown the possibility of instability without the need for external forcing. There are four possible types of instability. First, the energy mode, due to instability in the energy equation alone; this mode is most unstable at wavenumber $m=0$, leading to growth in energy across the whole domain. Next, there are three possible instabilities attributable to either the  Phillips effect or a $\gamma$-style instability, depending on the parameterisations of the fluxes. At low wavenumbers, there can be one or two positive growth rates, corresponding to positive eigenvalues of the Jacobian of the temperature and salinity fluxes with respect to their gradients. There is also the possibility of a high wavenumber instability, which can be avoided by suitable parameterisation of the flux terms.

We compare these results with those of two important previous studies. The system of \citet{BLY98} was unstable to a single Phillips mode, and also allowed for a stable energy mode. We have shown the possibility of an extra Phillips mode, allowing for an oscillatory instability, as may be expected in the diffusive convection regime. While BLY's use of the energy equation prevented the high wavenumber instability inherent to \citet{PHILLIPS197279} and \citet{Posmentier}, a careful choice of fluxes is necessary to avoid instability at high wavenumbers in our three-component system. The model of \citet{Radko_2003} also displays a high-wavenumber instability;  as in BLY, this is regularised by our inclusion of an energy equation. Further, we have demonstrated that Radko's $\gamma$-instability is mathematically equivalent to the Phillips effect in the case where turbulent fluxes depend on the density ratio alone. The multiscale analysis of \citet{radko2019thermohaline} provides an alternative method of regularising the high wavenumber instability by the inclusion of hyperdiffusion terms. This leads to a model that produces the key dynamics of double-diffusive layering in terms of only the temperature and salinity fields (in comparison with our three-component model). 
Radko's model depends on the empirical calibration of several coefficients, requiring direct numerical simulations \textit{a priori} to inform the choices of coefficients. By contrast, our model is derived using a spatial averaging process and the choice of a simple mixing length dependent on only one parameter. 
Both our model and that of \citet{radko2019thermohaline} provide similar numerical results, with staircases appearing and gradually reducing via the B-merger pattern of \citet{radko2007mechanics}.

We have applied the general results for the linear stability  of three-component systems in \S\,\ref{sec:3Phillips} to the model for double-diffusive convection presented in \S\,\ref{sec:model}. Our model describes the evolution of horizontally averaged temperature, salinity and turbulent kinetic energy fields, based on turbulent flux terms expressed in terms of a mixing length parameterised as a function of the dependent variables. There is no externally imposed stirring or energy source, in contrast to the model of \citet{PvH}, which also employed a BLY-like framework. The parameterisation of the length scale is a key ingredient of the model, and the form chosen~\eqref{eqn:ddclength} is not the only possibility. It is interesting to note that neither the length scale adopted by \citet{BLY98} and  \citet{Pruzina2022stirred} in a model of stirred convection, nor any similar simple length scale parameterised in terms of $b_z$, provides the appropriate release of potential energy from the flux terms to generate layering in (non-stirred) DDC. A prescription based on the density ratio $R$ alone leads to a high wavenumber instability. However, a relatively simple parameterisation in terms of both $R$ and $e$, with a form chosen based on the $\gamma$-instability theory, captures the essential physics of the layering process.

A slightly simpler mathematical model of layering could be produced by a reduced form of the model in which $e_t\equiv0$.
In this case, the energy equation becomes diagnostic and $e$ is determined directly at all times as a functional dependent on the global temperature and salinity fields. This modification does not affect the conditions for the layering instability \eqref{eqn:generalcond2}, but removes the possibility of the energy mode instability \eqref{eqn:generalcond1}. In this case, we are effectively left with a two-component system for $T$ and $S$, closed via a functional parameterisation of $e$ in the form of the steady energy equation. The layering instability, and its regularisation, are dependent entirely on the specific forms of $f$, $c$ and $p$.

In \S\,\ref{sec:stability}, we showed that the model~\eqref{eqn:ddcT}--\eqref{eqn:ddce} admits uniform-gradient steady states $(T_z,S_z,e) = (z,z/R_0,e_0)$ in the salt fingering regime, with $e_0(R_0)$ being single-valued for appropriate choices of $\delta$ and $\epsilon$. These steady states are unstable to perturbations for a finite range of $R_0$, with a well-defined wavenumber of maximum growth rate. Increasing the value of $\tau$ decreases the range of $R_0$ susceptible to instability, with no instability possible for $\tau \gtrsim 0.105$ (a value dependent on $\delta$ and $\epsilon$). Larger values of $\sigma$ give wider ranges of $R_0$ for instability. For values of the Prandtl number  $\sigma\lesssim 1$, the unstable range of $R_0$ is very narrow, and exists only for $R_0<1$, i.e. outside of the salt fingering regime. Hence, for sufficiently small Prandtl number, there is no possibility for salt fingering staircases to form via the $\gamma$-style instability, in agreement with the results of  three-dimensional numerical simulations of the full thermohaline system \citep{traxler2011numerically}.

We have presented numerical solutions to the model in \S\,\ref{sec:numerics}, showing the initial development of a staircase and its subsequent evolution to late times. The staircase evolves via the B-merger process described by \citet{radko2007mechanics}, eventually leaving a single strong interface with well-mixed layers on either side. During a merger event, the buoyancy gradient in the remaining interface increases, so that the gradient in the final interface is signficantly higher than in the first interfaces to develop. The buoyancy flux through the system also shows sharp increases during merger events, which can be explained by the condition responsible for the B-merger that flux increases as buoyancy jump decreases. This increase in flux agrees with previous results \citep[e.g.][]{Rosenblum_et_al_2011}, and is an important piece in the puzzle needed to understand transport processes through staircases.

One  important direction for further study is a comparison of this model with direct numerical simulations. The model may be tuned simply by choice of parameters, but also by reconsidering some of the assumptions on scalings in the original derivation. Through such fine-tuning, we hope to produce a more quantitative model that can be used to make accurate predictions for real staircase structures.

We saw in \S\,\ref{sec:stability} that the model in its current form does not admit steady states in the diffusive convection regime. Furthermore, numerical solutions of our model in this regime do not produce layers. It is therefore of significant interest to seek to adapt the model to  diffusive convection. \citet{ma2022thermohaline} suggest that an energy source is necessary for diffusive layering to occur in a reduced model, with double-diffusive effects acting to regularise the layering process, rather than being the driving factor behind it. Adding an energy source to~\eqref{eqn:ddce} does indeed allow the development of layers. However, three-dimensional numerical simulations \citep[e.g.][]{Stellmach_et_al_2011} have shown the formation of clear layers with no such source term. As such, further study is required to understand the diffusive convective regime using a mixing-length framework.

\section*{Acknowledgements}
PP is supported by the Natural Environment Research Council Panorama DTP [Grant No. NE/S007458/1]. We are grateful to participants at the KITP Staircase 21 Programme, [National Science Foundation  Grant No. NSF PHY-1748958] for helpful discussions, and to three anonymous reviewers for a number of insightful comments that have helped improve the paper.
\section*{Declaration of interests}
The authors report no conflict of interest

\appendix
\section{Derivation of the model}\label{sec:modelderivation}
This appendix contains a review of the derivation of the system~\eqref{eqn:ddcT}--\eqref{eqn:ddce} from the Boussinesq equations \eqref{eqn:boussunondim}--\eqref{eqn:boussstatenondim} using the horizontal averaging process of \citet{Pruzina2022stirred}. 

We define the horizontal average of a quantity $q(\bm{x},t)$ over a spatial area $A$ at a height $z$ as
\begin{equation}
\langle q \rangle \equiv \frac{1}{A}\int_A q(\bm x,t) \; \textrm{d}A. \label{eqn:averaging}
\end{equation}
Let $\bm{u}_h = (u,v)$ represent the horizontal velocity and $\nablab_h$ the horizontal gradient operator. The variables will be considered in terms of the sum of their horizontal mean and fluctuation components: $T = \langle T\rangle + T'$, $S = \langle S\rangle + S'$, $\bm{u}_h= \langle \bm{u}_h\rangle + \bm{u}_h'$, ${w}= \langle w\rangle + {w}'$. 

Taking the average of the incompressibility condition~\eqref{eqn:boussincompnondim} and applying either the condition of impermeability or periodicity on the horizontal components of the velocity, we obtain
\begin{equation}
\langle \nablab_h\cdotb\bm{u}_h\rangle + \langle w_z\rangle = 0, \quad \textrm{and hence} \quad  \langle w\rangle_z= 0.\label{eqn:wzzero}
\end{equation}
Thus, the horizontally averaged vertical velocity is uniform across the height of the domain. Assuming impermeability conditions on the top and bottom boundaries, it follows that $\langle w\rangle = 0$. Thus there is no mean vertical velocity and $w=w'$.

We demonstrate the averaging process for the temperature equation. Beginning with~\eqref{eqn:boussTnondim}, we take the horizontal average and apply~\eqref{eqn:wzzero} to give
\begin{equation}
\langle T\rangle_t + \langle wT'\rangle_z = \langle T\rangle_{zz}.\label{eqn:boussTmean}
\end{equation}
We subtract~\eqref{eqn:boussTmean} from the full temperature equation~\eqref{eqn:boussTnondim}, and apply a quasilinear approximation to neglect both the perturbation-perturbation term $\bm{u}'_h\cdotb\nablab_hT'$ and the perturbation temperature flux $wT'_z-\langle wT'\rangle_z$, giving
\begin{equation}
T'_t + w\langle T\rangle_z = \nabla^2T'.\label{eqn:boussTfluctuation}
\end{equation}
We now use a scaling argument to parameterise the term $\langle wT'\rangle$ in terms of mean quantities. We assume that the mean of the square of the vertical velocity is a constant multiple of the mean total kinetic energy $\langle e\rangle = \langle\bm{u}\cdotb\bm{u}\rangle/2$, i.e.
\begin{equation}
\langle w^2\rangle = \beta^2\langle\bm{u}\cdotb\bm{u}\rangle/2 = \beta^2\langle e\rangle,\label{eqn:energyassumption}
\end{equation}
for some dimensionless constant $\beta$. While, in general, the value of $\beta$ may vary, we assume it is constant here to avoid overcomplicating the system. We assume that the turbulence varies on a mixing length scale $l$, and on the dynamical timescale $\tau_d\sim l/\beta\langle e\rangle^{1/2}$, i.e. the characteristic time to move a vertical distance $l$. The length scale will be parameterised later in terms of the dependent variables, as discussed in \S\ref{sec:model}. With these assumptions, we approximate the time and space derivatives as $\partial_t \sim 1/\tau_d = \beta\langle e\rangle^{1/2}/l$ and $\nabla^2\sim-1/l^2$. Multiplying \eqref{eqn:boussTfluctuation} by $w$, averaging, using these scalings, rearranging, and applying \eqref{eqn:energyassumption}, gives the turbulent temperature flux as
\begin{equation}
    \langle wT'\rangle = -\beta^2\frac{l^2\langle e\rangle}{\beta l\langle e\rangle^{1/2} + 1}\langle T\rangle_z.\label{eqn:tempflux}
\end{equation}
Combining~\eqref{eqn:boussTmean} and~\eqref{eqn:tempflux} then gives the mean temperature equation as
\begin{equation}
\frac{1}{\beta}\langle T\rangle_t = \left(\frac{l^2\langle e\rangle}{l\langle e\rangle^{1/2} + \beta^{-1}}\langle T\rangle_z\right)_z + \frac{1}{\beta}\langle T\rangle_{zz}.
\end{equation}
In this form, the dimensionless parameter $\beta$ acts simply as a scale factor on the time derivative, effectively setting a new dimensionless time variable $\tilde{t} = \beta t$. This can be incorporated into the nondimensionalisation using a scaled thermal diffusivity $\tilde{\kappa}_T = \beta\kappa_T$. With this transformation, and dropping angled brackets, the temperature equation becomes
\begin{equation}
 T_t = \left(\frac{l^2 e}{l e^{1/2} + 1} T_z\right)_z,
\end{equation}
where we assume that the molecular diffusion term is small and hence can be neglected. The salinity equation~\eqref{eqn:ddcS} is obtained by exactly the same process. To obtain an equation for the mean kinetic energy, we begin by formulating the three-dimensional energy equation by taking the scalar product of the momentum equation~\eqref{eqn:boussunondim} with $\bm{u}$, giving
\begin{equation}
e_t + \nablab\cdotb\left(\bm{u}e\right) = -\sigma\nablab\cdotb\left(\bm{u}p\right) + wb + \sigma\left(\nabla^2e - |\nablab \bm{u}|^2\right).\label{eqn:bousse}
\end{equation}
The terms on the right-hand side represent, in order, the effect of pressure, conversion from potential to kinetic energy, diffusion of kinetic energy, and the viscous dissipation $D = -\sigma|\nablab\bm{u}|^2$.
The horizontal average of~\eqref{eqn:bousse} gives
\begin{equation}
\langle e\rangle_t + \langle we'\rangle_z - \sigma\langle e\rangle_{zz} = \langle wb'\rangle - \langle D\rangle. \label{eqn:boussemean}
\end{equation}
The turbulent energy flux $\langle we'\rangle$ is parameterised using the same method as for the temperature flux; the buoyancy flux is calculated as the temperature flux minus the salinity flux. Following BLY, the dissipation term is parameterised by $D = \epsilon e^{3/2}/l$. This form is commonly used in $k$-$\epsilon$ models of turbulence \citep[e.g.][]{jones1972prediction}. We finally obtain the energy equation
\begin{equation}
e_t = \left(\frac{l^2e}{le^{1/2} + \sigma}e_z\right)_z - \sigma\left(\frac{l^2e}{le^{1/2}+1}T_z - \frac{l^2e}{le^{1/2}+\tau}S_z\right) + \sigma e_{zz} - \epsilon \frac{e^{3/2}}{l},
\end{equation}
leading to the full system~\eqref{eqn:ddcT}--\eqref{eqn:ddce}.

In the derivation above, we have assumed that $\nabla^2\sim-1/l^2$ for each of the temperature, salinity and kinetic energy fields. A more accurate representation could possibly be obtained by using different scalings in each equation to reflect the different molecular diffusivities of each field. Here, however, we use the same scalings to produce a relatively simple model that admits staircase solutions.

\section{Merger behaviour}\label{sec:oscillatorderivation}
This appendix contains the derivation of the nonlinear oscillator equation~\eqref{eqn:nussoscillator} for the energy, by considering constant flux steady states. The process outlined below mirrors that of \citet{BLY98} for their two-component system. Our derivation here includes the third equation in our three-component model.

Assuming a steady state, the system~\eqref{eqn:ddcT}--\eqref{eqn:ddce} can be written as
\begin{eqnarray}
f_0 = \frac{D^2}{D+1}T_z,\label{eqn:nussT}\\
c_0 = \frac{D^2}{D+\tau}S_z,\label{eqn:nussS}\\
0 = \left(\kappa e_z\right)_z -\sigma\left(f_0-c_0\right) - \epsilon\frac{e^2}{D},
\label{eqn:nusse}
\end{eqnarray}
where $D = le^{1/2}$ and $\kappa = \left(D^2/\left(D+\sigma\right) + \sigma\right)$. 
Note that~\eqref{eqn:nussT} uniquely defines $D(T_z)$ and~\eqref{eqn:nussS} uniquely defines $S_z(D)$. The salt gradient is therefore tied to the temperature gradient, rather than the two fields being independent.

Dividing~\eqref{eqn:nussT} by~\eqref{eqn:nussS}, we obtain
\begin{equation}
\gamma_0 = \frac{f_0}{c_0} = \frac{D+\tau}{D+1}R,\label{eqn:nussR}
\end{equation}
thus defining $R$ in terms of $D$, where $\gamma_0 = f_0/c_0$ is the steady-state flux ratio.
The prescription for the length scale~\eqref{eqn:ddclength}, coupled with the definition of $D = le^{1/2}$, can be rearranged to give an expression for $e(D,R)$, namely,
\begin{equation}
e^2 = \left(D^2-\delta\right)R^2.
\label{eq:Appx_eR}
\end{equation}
Combining \eqref{eq:Appx_eR} with \eqref{eqn:nussR}, we define the energy solely in terms of $D$ by
\begin{equation}
e(D) = \frac{\sqrt{D^2-\delta}\left(D+1\right)}{D+\tau}\gamma_0.\label{eqn:nusse(D)2}
\end{equation}
Multiplying the energy equation~\eqref{eqn:nusse} by $\kappa e_z$, integrating with respect to $z$, and changing variables such that $e_z  \intd z = e_D \intd D$, we obtain
\begin{equation}
\frac{1}{2}\left(\kappa e_z\right)^2 + \bigintsss \left( -\sigma\left(f_0-c_0\right) - \epsilon\frac{e(D)^2}{D} \right)\kappa e_D \intd D = E,
\label{eqn:nusspoteq}
\end{equation}
where $E$ is a constant.

To transform~\eqref{eqn:nusspoteq} to an equation for $e_b$, we could write $e_z = e_bb_z$ and divide all terms in~\eqref{eqn:nusspoteq}  by $(\kappa b_z)^2$. However, when $b_z = 0$, this leads to a singularity in the potential. To avoid this, we note that $T_z\neq 0$ when $f_0\neq0$, and write $e_z = e_T T_z$ in terms of the temperature instead. This use of $T$ instead of $b$ as the `time' variable is valid because $D$ and $S_z$ are defined as functions of $T_z$ via~\eqref{eqn:nussT}--\eqref{eqn:nussS}. On using \eqref{eqn:nussT} to rewrite $T_z$ as a function of $D$, we obtain
\begin{equation}
\frac{1}{2}e_T^2 + U(D) = E\left(\frac{ D^2}{\kappa f_0 \left(D+1\right)}\right)^2,\label{eqn:nussoscillator2}
\end{equation}
where the potential $U(D)$ is defined by
\begin{equation}
U(D) = \left(\frac{D^2}{\kappa f_0 \left(D+1\right)}\right)^2 \bigintsss \left( -\sigma\left(f_0-c_0\right) - \epsilon\frac{e(D)^2}{D} \right)\kappa \intd  D.\label{eqn:nusspotential2}
\end{equation}

\bibliographystyle{jfm}
\bibliography{bibliography}

\end{document}